\documentclass[12pt,preprint]{aastex}

\shorttitle{Interferometry of Transitional Disks}
\shortauthors{Pott et al.}

\begin{document}


\title{Ruling out Stellar Companions and Resolving the Innermost Regions of Transitional Disks with the
	Keck Interferometer
}


\author{Jorg-Uwe~Pott\altaffilmark{1,2,3}, Marshall~D.~Perrin\altaffilmark{1}, Elise~Furlan\altaffilmark{1,4}, Andrea~M.~Ghez\altaffilmark{1,5}, Tom~M.~Herbst\altaffilmark{2}, Stanimir~Metchev\altaffilmark{1,6}}

\altaffiltext{1}{Div. of Astronomy \& Astrophysics, University
of California, Los Angeles,  CA 90095-1547\newline {\tt mperrin@astro.ucla.edu, ghez@astro.ucla.edu}}
\altaffiltext{2}{current address: Max-Planck-Institut f{\" u}r Astronomie, K{\" o}nigstuhl 17, D-69117 Heidelberg, Germany\newline {\tt jpott@mpia.de, herbst@mpia.de}}
\altaffiltext{3}{W.~M.~Keck~Observatory, California Association for Research in Astronomy, Kamuela, HI 96743}
\altaffiltext{4}{current address: Jet Propulsion Laboratory, Caltech, 4800 Oak Grove
 Drive, Pasadena, CA 91109\newline {\tt elise.furlan@jpl.nasa.gov}}
\altaffiltext{5}{Institute of Geophysics and Planetary Physics,
	University of California, Los Angeles, CA 90095-1565
}
\altaffiltext{6}{current address: Dept. of Physics and Astronomy, State University of New York,
Stony Brook, NY 11794-3800\newline {\tt metchev@astro.sunysb.edu}}

\begin{abstract}
{  With the Keck Interferometer, we have studied at 2~um the innermost regions  
of several nearby, young, dust depleted 		
	"transitional" disks.  Our observations target five of the six  
clearest cases of		transitional disks in the Taurus/Auriga star-forming region (DM Tau, GM Aur, LkCa 15, UX Tau A, and
	RY Tau) to explore the possibility that the depletion of optically thick dust from the
inner disks is caused by stellar companions rather than the more typical planet-formation
hypothesis. 
		At the 99.7\%  
confidence level, the observed visibilities
	exclude binaries with flux ratios of at least 0.05 and separations ranging from 2.5 to 30 mas (0.35 - 4 AU) over $\gtrsim\,$94\% of the area covered by our
	measurements.  All targets but DM~Tau show near-infrared excess in their SED higher than our companion flux ratio detection limits.   While a companion has previously been detected in  
the candidate transitional disk system \object{CoKu~Tau/4},  we can exclude similar mass companions as the typical origin for
	the clearing of inner dust in transitional disks and of the near-infrared excess emission.  Unlike CoKu~Tau/4, all our targets show some evidence of accretion.   
We find that all but one of the
	targets are clearly spatially resolved, and UX~Tau~A is marginally resolved.
	Our data is consistent  
with hot material on small scales (0.1 AU)  
inside of and separated from the cooler outer disk, consistent  
with the recent SED modeling.  
These observations support the notion 		that some transitional disks  
have radial gaps in their optically thick material, which could be an 		
	indication for planet formation in the habitable zone ($\sim$ a few
AU) of a protoplanetary disk.
}
\end{abstract}




\section{Introduction}

Circumstellar disks are a natural outcome of the star-formation
process: when a molecular cloud core collapses, it gives rise to a central star surrounded by a rotating circumstellar
disk, which transports material towards the star. Over time, the disk material dissipates
through {  processes such as} accretion onto the central star,  {  disk winds} and the formation of
planets. At an age of $\sim5~{\rm Myr}$, about 90\% of disks have already dispersed, and within $10~{\rm Myr}$ of their formation, almost all pre-main-sequence stars are diskless \citep[e.g.][]{2006ApJ...638..897S}. 
While it is now believed that such disks commonly give rise to planetary systems,
		the details of this process remain unclear.
Theory predicts that disks evolve from the inside out: dust grain growth is expected to
occur faster in the inner disk than in the outer disk \citep[e.g.][]{2004Natur.432..479V,2005A&A...434..971D}, higher densities favor planet formation in the inner disk \citep{2002Icar..156..291B}, and photoevaporation by the central star will cause the inner disk to dissipate first \citep{2001MNRAS.328..485C}. 

{  Possible observational support for inside out disk evolution has been found in a
		small number of so-called transitional disks.}
These systems show a strong mid-infrared excess ($\gtrsim\,8\,\mu {\rm m}$) revealing the presence of dust {  but significantly reduced or
no shorter wavelength infrared excess compared to typical classical T Tauri disks, indicating a depletion of optically thick
inner dust out to a radius of a few AU}
\citep{1990AJ.....99.1187S, 1997ApJ...489L.173M, 2001AJ....121.1003S, 2005ApJ...621..461D, 2005ApJ...630L.185C, 2006ApJ...636..932M,
2006ApJS..165..568F, 2006AJ....131.1574L, 2006ApJ...643.1003M, 2007ApJ...664L.111E_disk, 2007ApJ...664L.107B}. 
Therefore, these disks might be in the process of dispersing
and this has often been assumed to be due to the influence of newly formed planets \citep[e.g.][]{2004ApJ...612L.137Q}.  
{  Discussed explanations of the transitional disk phenomenon reveal two important features which can be tested directly by high angular resolution imaging observations.

(1) The depletion of dust inside of the outer, mid-infrared disk, could be caused by a close (AU-scale) binary system inside of the disk.} {  Binary companions can perturb a circumstellar disk and create inner holes with diameters comparable to the binary separation \citep[][see also the discussion for DI~Tau in Meyer et al. 1997]{1994ApJ...421..651A}.} 
To call such a circumbinary disk
{\it transitional}, would be misleading, since circumbinary disks can be dynamically stable and longer-lasting than the short ($<$~Myr) time-scales derived from the small (few percent) fractional abundance of transitional disks around pre-main-sequence stars in nearby, a few Myr young, star-forming regions \citep[][note that the fractional abundancies of transitional disks might be as high as a few tens of percent, depending on the exact definition of transitional disks, in particular if and what type of {  residual} inner disk  emission is {  permitted}]{2006ApJS..165..568F,2008AJ....135..966F,2009ApJ...703.1964F}. 
{  Also, a close companion affects the
SED interpretation of apparent transitional disk systems.} Unresolved infrared
companions can create additional near-infrared and, if embedded, mid-infrared flux, that appears comparable to the infrared excess radiation seen in transitional disk systems, which typically is interpreted as disk emission \citep[e.g.][]{2003ApJ...592..288D}. 
{  Indeed, recent diffraction limited NIR imaging with the Keck II telescope of the candidate transitional disk system \object{CoKu~Tau/4} indicates that its inner hole (10~AU radius) is actually caused by a newly discovered binary companion of $\sim\,8\,{\rm AU}$ separation, removing the need to invoke other processes like planet formation as the disk clearing mechanism in transitional disk systems \citep{2008ApJ...678L..59I}.}
While the census of very close 
companions {  of T~Tauri stars (TTS) in star-forming regions} is far from complete,
the companion star fraction in young, nearby star-forming regions is about 50\% in the 15-1800~AU separation range and $\sim\,20\%$ at separations less than 10~AU \citep{1993AJ....106.2005G,1997ApJ...481..378G,1993A&A...278..129L,1995ApJ...443..625S}. The companion star fraction typically
decreases towards smaller separations (less than a few AU), but
\citet{2003A&A...410..269M} suggest that YSOs in Ophiuchus 
have a companion fraction of at least 10\% at the 0.8-4~AU separation scale. 
{  These observational constraints suggest that there would be enough binaries to populate a large fraction of transitional disks, although the short-period binary frequency appears to vary between different sites of star formation \citep{1992ASPC...32...30M, 2003A&A...410..269M}. } 

{  (2) While some transitional disks may be completely cleared of material in the inner region,
the planet formation hypothesis suggests that disk clearing may often result in gaps between inner and outer dusty
regions \citep{2004ApJ...612L.137Q,2004A&A...425L...9P,2006ApJ...640.1110V}. Other possible disk clearing mechanisms such as photoevaporation
would produce strictly inside-out clearing \citep{2000prpl.conf..401H, 2001MNRAS.328..485C}, so evidence for
gaps in disks (in contrast to totally cleared holes) tends to support the planet formation hypothesis.
Many transitional disks show some infrared excess emission inside of the outer optically thick disk, which itself dominates at wavelengths longer than $\sim\,8~\mu$m.
It has been shown for a few systems that this near-infrared excess can be explained by a small amount of emitting dust close to the star at $\sim\,0.1\,$AU-scales, leaving a gap between this innermost dust, and the outer mid-infrared disk \citep[e.g.][]{2008ApJ...682L.125E_LkCa}.
Therefore, the near-infrared excess in transitional disk systems might origin from such small size scales, if not emitted by a so far unresolved companion (case 1).}

To {  directly} assess (1) the presence of close binary companions within transitional disks and {  (2) the emission size scale of the near-infrared excess over the stellar continuum}, we used the Keck Interferometer (KI)
in $V^2$ mode\footnote{$V^2$-mode measures the continuum squared visibility amplitude of the astronomical $K'$-band} to observe 5 transitional disks in the nearby ($\sim$~140~pc) Taurus-Auriga young
star-forming {  region}. 
The nominal interferometric resolution of $\sim\,2.7$~mas and the field of view of $\sim\,50\,{\rm mas}$, offered in the $V^2$ mode, is well suited to resolve any companion stars from about 0.5 to 5~AU {  distance} from the target primary stars. This angular resolution is a significant improvement over the resolution
available with speckle or aperture mask interferometry and adaptive optics at 8-10~m class telescopes ($\gtrsim$~25~mas). 

This article is organized as follows: {  detailed target properties }are reported in Sect.~\ref{sec:2}.
Observations and data reduction are give in Sect.~\ref{sec:3}.
The results are discussed in Sect.~\ref{sec:5}, and the conclusions of our experiment are given in Sect.~\ref{sec:6}

\section{\label{sec:2}Target selection and properties}

\begin{table}
\begin{center}
\caption{\label{tab:1}
Stellar properties of the targets, and disk properties from previously published models. 
}
\begin{tabular}{lccccc|c}
\tableline\tableline
    & DM~Tau$^a$ & GM~Aur$^a$ & LkCa~15$^b$ & UX~Tau~A$^b$ & RY~Tau$^c$ & CoKu~Tau/4$^f$\\
\tableline
$M_* \,({\rm M_{\odot}})$ & 0.65 & 1.2 & 1.1 & 1.5 & 2.0 & 0.5 \\
$R_* \,({\rm R_{\odot}})$ & 1.2  & 1.5 & 1.7 & 2.0 & 3.6 & 1.9$\,^{h}$ \\
$T_* \,({\rm K)}$          & 3720& 4730 & 4350 &4900& 5945$\,^{d}$ & 3720\\
$L_* \,({\rm L_{\odot}})$ & 0.25 &1.03 &0.96 &2.18 & 12.8 & 0.61$\,^{h}$ \\
$\dot{M} \, ({\rm M_{\odot}\,yr^{-1}})$& 2.0$\cdot 10^{-9}$ & 1.0$\cdot 10^{-8}$& 2.4$\cdot 10^{-9}$ & 9.6$\cdot 10^{-9}$ &  2.5$\cdot 10^{-7}$ & --\\
Spectral type & M1 & K5 & K5 & K2 & G1$\,^{d}$ & M1.5 \\
$A_{\rm V}$   & 0.5 & 1.2 & 1.2 & 1.3 & 2.1 & 3.0~$^g$ \\
$K_{\rm s}$ mag.~$^e$ & 9.5 & 8.3 &  8.2 & 7.5 & 5.4 & 8.3~$^g$ \\
\tableline
Inclination (deg)    & 40 & 55 & 42 & 60 & 25 & $50\,-\,75\,^{i}$\\
$R_{\rm hole}\,^l\,{\rm (AU)}$ & 3 & 24 & 46 &  56 & $^{k}$ & 10\\
$K_{\rm inner\,dust}/K_{\rm total }$ & $\lesssim\,0.05$ & 0.12 & 0.23 & 0.32 & 0.73 &$\lesssim\,0.05$\\
$R_{\rm inner\,dust}\,^l\,{\rm (AU)}$ & -- &$<\,5$& $0.13$ & $0.17$ & $0.25$ & --\\
\tableline
\end{tabular}
\tablenotetext{a}{\citet{2005ApJ...630L.185C}}
\tablenotetext{b}{\citet{2007ApJ...670L.135E_LkUX}}
\tablenotetext{c}{\citet{2005ApJ...622..440A_PTI}}
\tablenotetext{d}{\citet{2004AJ....128.1294C}}
\tablenotetext{e}{\citet{2006AJ....131.1163S}}
\tablenotetext{f}{{  To put our five target stars in context with CoKu~Tau/4, the respective values are given based on the disk model presented by \citet{2005ApJ...621..461D}.}}
\tablenotetext{g}{{  This relatively large $A_{\rm V}$ is attributed to foreground extinction outside the CoKu~Tau/4 system. Therefore we report here the extinction-corrected $K$-band magnitude.}}
\tablenotetext{h}{{  Note that the radius and luminosity discrepancy between CoKu~Tau/4 and DM~Tau despite of their similar spectral type is easily explained by the recently detected, close equal-mass companion inside the disk hole \citep{2008ApJ...678L..59I}.}}
\tablenotetext{i}{{  \citet{2005ApJ...621..461D} successfully fitted models within this range of inclinations. Furthermore, they argue against edge-on geometry based on the stellar luminosity. This argument is corroborated by the detection of the visual binary with regular luminosities \citep{2008ApJ...678L..59I}}}
\tablenotetext{k}{  No published model of the disk hole size.}
\tablenotetext{l}{  These two radii give characteristic sizes of the disk SED models. $R_{\rm inner\,dust}$ refers to the modeled size scale of the near-infrared emission inside $R_{\rm hole}$. The emission from $R_{\rm inner\,dust}$-scales is believed to be responsible for the NIR excess $K_{\rm inner\,dust}/K_{\rm total }$. $R_{\rm hole}$ instead scales the inner edge of the outer mid-infrared dominating disk. See Fig.~4 in 
\citet{2008ApJ...682L.125E_LkCa} for a sketch.}
\tablecomments{ The fractional $K$-band excess $K_{\rm inner~dust}/K_{\rm total}$ is the flux ratio of
disk to total light at $K$ band, as inferred from previous models. 
Those models were based on spatially unresolved photometry, and, in the case of RY~Tau, also on interferometric data from PTI.  Note that \citet{2005ApJ...635.1173A_KI} report a fractional $K$-band excess of 0.43 for LkCa~15 from pure photometry fits. Here we use the 0.23 from \citet{2007ApJ...670L.135E_LkUX} however, since they included NIR spectra in the fit, which clearly show and quantify the hot dust emission.
}
\end{center}
\end{table}

{  Transitional disks are defined as systems that are significantly depleted of optically thick dust on scales of a few AU compared to the majority of similarly aged stars with circumstellar disks.}
However, the variety of observing constraints {  and their interpretation as well as their possible dependence on age and environment makes it difficult to describe a {\it typical} transitional disk}. 
Some disk data favor an inner {\it hole}, which refers to a true depletion of the inner optically thick dust \citep{2007ApJ...664L.111E_disk}.
A disk {\it gap} is present, when optically thick dust close to the stellar photosphere, as traced by 2~$\mu$m excess emission, is separated from the outer, cooler, optically thick dust disk, which is detected at mid-to-far-infrared and ${\rm (sub-)mm}$ wavelengths \citep{2008ApJ...682L.125E_LkCa}. 
In general, it is assumed that dust (traced by the IR continuum) and gas (traced by emission lines) are well mixed in primordial disks. 
In individual systems however, gas has been found in disk regions devoid of optically thick dust \citep[e.g.][]{2008ApJ...675L.109B, 2009ApJ...699..330S}. Thus, an observed depletion of optically thick dust does not need to correlate with a similarly complete depletion of disk material. 

Our sample consists of five low-mass pre-main sequence stars in the Taurus-Auriga star forming region showing mid-infrared dust excess characteristic of transitional disks: \object{DM Tau}, \object{GM Aur}, \object{LkCa 15}, \object{UX Tau A}, and \object{RY Tau}. 
These objects comprise the best-studied disks to date (i) with strong evidence from observations or SED-modeling that inner disk holes or gaps are present \citep[][
see Table~\ref{tab:1}]{2005ApJ...621..461D,2005ApJ...630L.185C, 2007ApJ...670L.135E_LkUX} and (ii) for which no close binary companions are known (i.e., all but CoKu~Tau/4). 

Table~\ref{tab:1} lists the stellar and disk properties of our targets and associated references. The respective values of CoKu~Tau/4 are given as well for comparison. 
  DM~Tau is the only object in our sample with no detectable excess
  emission below 8~$\mu$m, but it is still accreting, traced by hydrogen emission lines.
  This suggests that
  at least gas must exist in the innermost disk, feeding the accretion. GM~Aur has a hole of 24~AU, which is partially filled with optically thin dust. Both LkCa~15
  and UX~Tau~A seem to have gaps between optically thick inner
  and outer disk regions; in addition, LkCa~15 also has some optically
  thin dust in the gap. RY~Tau could be a transitional disk with a gap, in part
  due to its somewhat similar SED shape to LkCa~15  \citep{2009ApJ...703.1964F}. On the other hand,
  RY Tau has an earlier spectral type than the rest of the sample, and
  therefore its disk structure could be different.

Our data probe binary separations of 2.5-30~mas (0.35-4~AU). This matches the inner region, which cannot be resolved by single telescope imaging of 8-10~m class telescopes, and which lies inside the colder optically thick dust disks, responsible for the MIR-excess of transitional disks (see $R_{\rm hole}$ in Table~\ref{tab:1}).
{  In particular, this region covers the range where a stellar companion might reside and be responsible for inner disk truncation or NIR excess flux.}
A circumbinary disk is typically truncated at an inner radius of about 2-3 times the semi-major
axis of the binary.
 \citet{1994ApJ...421..651A} investigated the gravitational impact of a central binary on a geometrically thin, non-self gravitating circumbinary disk, and found that the semi-major axis of the central binary is about half the inner disk edge-radius. The detailed relation between binary separation and disk hole size depends on the binary's eccentricity.

{  Our interferometric approach} is sensitive to companions
with orbital periods ranging from several months to several years, a regime that could also
be probed by radial velocity surveys.
However, such surveys would take years to complete, as opposed to a single epoch of KI
observations. Earth's
rotation of the KI baseline allows us to obtain visibilities at about 5 independent {\it spatial frequencies} (or {\it  u,v}-points) per object per night, {  which suffices to exclude a large number of binary parameters.}
 For some of our targets, we are able to combine
our data with previous interferometric measurements; RY Tau: \citet{2005ApJ...622..440A_PTI}, LkCa~15, GM~Aur: \citet{2005ApJ...635.1173A_KI}.  
{  The earlier measurements of these objects' visibility amplitudes indicate that they are spatially resolved and were modeled as inner disk structure in each case. However, those single visibility measurements of LkCa~15 and GM~Aur are insufficient to distinguish such disk structure from a companion star.   This ambiguity is easily resolved through the visibility amplitude measurements at multiple spatial frequencies presented here.}

\section{Observations and data reduction}
\label{sec:3}
\begin{figure}
\epsscale{.80}
\plotone{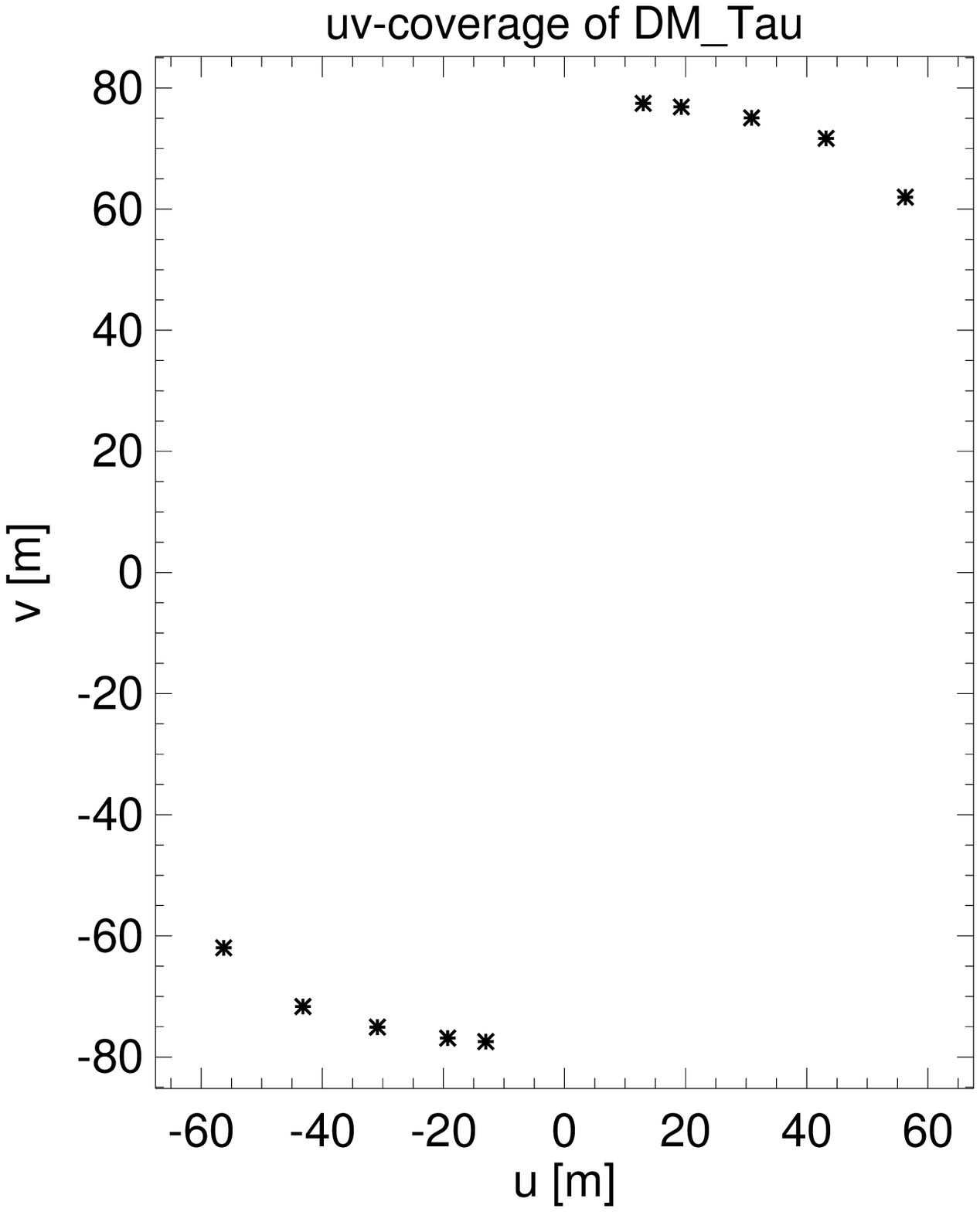}
\caption{\label{fig:1}
{\it u,v}-coverage of our KI observations of DM~Tau. A significant {\it u,v}-coverage is necessary to probe a wide range of binary parameters. This {\it u,v}-coverage is typical for the other targets, since all have similar declinations as DM~Tau. Note that each measurement creates two symmetric points in the {\it u,v}-plane.}
\end{figure}

\begin{table}
\begin{center}
\caption{\label{tab:4}
Observing log.}
\begin{tabular}{ccccccc}
\tableline\tableline
Target & date (UT) & H.A.$\,^{a}$ & $u,v$$\,^{b}$ & proj. B$\,^{b}$ & calibrators & $V^2$ \\
 && $[$hr$]$ &$[{\rm m}]$ & $[$m,\,deg(EofN)$]$  & [from Tab~\ref{tab:3}]& [calib.] \\
\tableline
DM~Tau & Mar.17,2008 &  3.3 & (16.7, 77.1) & (78.9, 12.2) & 1,2 & 0.91 \\
       & Mar.17,2008 &  3.4 & (15.3, 77.3) & (78.8, 11.2) & 1,2 & 0.91 \\
       & Dec.15,2008 & -1.2 & (56.3, 62.0) & (83.7, 42.2) & 6,7 & 0.90 \\
       & Dec.15,2008 &  1.1 & (43.2, 71.7) & (83.7, 31.1) & 6,7 & 0.90 \\
       & Dec.15,2008 &  2.2 & (30.9, 75.1) & (81.2, 22.4) & 6,7 & 0.88 \\
       & Dec.15,2008 &  3.1 & (19.3, 76.9) & (79.3, 14.1) & 6,7 & 0.90 \\
       & Dec.15,2008 &  3.5 & (13.0, 77.5) & (78.6, 9.5) & 6,7 & 0.89 \\
\tableline
GM~Aur & Mar.17,2008 &  2.6 & (25.7, 80.2) & (84.2, 17.8) & 1,2 & 0.95 \\
       & Dec.15,2008 & -0.2 & (52.7, 65.0) & (83.7, 39.0) & 6,7 & 0.93 \\
       & Dec.15,2008 &  1.7 & (36.9, 76.4) & (84.8, 25.8) & 6,7 & 0.95 \\
       & Dec.15,2008 &  3.4 & (14.2, 82.4) & (83.6, 9.8) & 6,7 & 0.96 \\
\tableline
LkCa~15 & Mar.17,2008 &  3.7 & (10.9, 79.8) & (80.5, 7.8) & 1,2 & 0.93 \\ 
        & Dec.15,2008 & -1.0 & (55.9, 62.2) & (83.6, 41.9) & 6,7 & 0.91 \\
        & Dec.15,2008 &  1.5 & (39.0, 74.2) & (83.8, 27.7) & 6,7 & 0.91 \\
        & Dec.15,2008 &  2.6 & (26.3, 77.7) & (82.0, 18.7) & 6,7 & 0.91 \\
        & Dec.15,2008 &  3.3 & (16.3, 79.3) & (81.0, 11.6) & 6,7 & 0.91 \\
\tableline
UX~Tau~A & Dec.15,2008 & -1.3 & (56.4, 61.7) & (83.6, 42.4) & 3,4,5 & 0.96 \\
         & Dec.15,2008 &  0.9 & (45.3, 70.8) & (84.1, 32.6) & 3,4,5 & 0.97 \\
         & Dec.15,2008 &  1.8 & (35.7, 74.0) & (82.2, 25.8) & 3,4,5 & 0.97 \\
         & Dec.15,2008 &  3.0 & (20.3, 76.8) & (79.4, 14.8) & 3,4,5 & 0.96 \\
\tableline
RY~Tau$\,^{d}$ & Mar.17,2008  & 2.6 & (25.6, 79.8) & (83.8, 17.8) & 1,2$\,^{c}$ & 0.30 \\
       & Dec.15,2008, & 1.5 & (39.4, 75.1) & (84.8, 27.7) & 3,4,5 & 0.27 \\
       & Dec.15,2008  & 2.7 & (24.4, 80.1) & (83.7, 16.9) & 3,4,5 & 0.28 \\
\tableline
\tableline
\end{tabular}
\tablenotetext{a}{hour angle}
\tablenotetext{b}{the {\it u,~v}-coordinates, given here in meters, are the baseline length (B) projected onto the line of sight. {\it u} points East, {\it v} points North. They are equivalent to polar values of the projected baseline, given in the next column in meters, and degrees East of North (compare to the {\it u,~v}-coverage plot in Fig.~\ref{fig:1}). Note that the spatial frequency, inversely related to the interferometric resolution, is given by $B/\lambda$, the ratio of the variable baseline projection and the fixed observing wavelength (2.2~$\mu$m in our case).}
\tablenotetext{c}{Note that these calibrators are, unlike during the Dec08 observations, significantly fainter than RY~Tau, which can increase the systematic error of the visibility calibration by a few percent for the KI. Without a bright calibrator measurement during that night, we cannot quantify this increased calibration uncertainty for that particular night, but since the raw visibility, i.e. the system transfer function, decreases with flux, this measurement appears to be very consistent with the other data (cf. Fig.~\ref{fig:2})}
\tablenotetext{d}{\label{foot:1}   The PTI visibilites from \citet{2005ApJ...622..440A_PTI}, used in our modeling of the $K$-band emission of RY~Tau, are $V^2(85m, 70^\circ)\,\sim\,0.33$, $V^2(109m, 17^\circ)\,\sim\,0.20$.}
\end{center}
\end{table}

\begin{table}
\begin{center}
\caption{\label{tab:3}
Properties of the interferometric calibrator stars used for calibration of the instrumental transfer function during data reduction. }
\begin{tabular}{clccc}
\tableline\tableline
\# &Calibrator & $V/H/K$ & Spec. Type$\,^{(a)}$ & Ang. diameter (mas)$\,^{(b)}$  \\
\tableline
1 & HD283798$^{(c)}$ & 9.5/8.1/8.0 & G2V   & $0.11\,\pm\,0.01$ \\
2 & HD283886         & 9.9/8.6/8.4 & G2V   & $0.08\,\pm\,0.01$ \\
3 & HD21379          & 6.3/6.3/6.3 & A0V   & $0.12\,\pm\,0.01$ \\
4 & HD254236         & 8.8/6.5/6.3 & K2III & $0.34\,\pm\,0.13$ \\
5 & HD41076          & 6.1/6.1/6.1 & A0V   & $0.13\,\pm\,0.01$ \\
6 & HD250388         & 10.7/8.9/8.8 & K0   & $0.09\,\pm\,0.01$ \\
7 & HD283934         & 10.6/9.0/9.0 & G5V  & $0.07\,\pm\,0.01$ \\
\tableline \tableline
\end{tabular}
\tablenotetext{a}{from {\sc Simbad}}
\tablenotetext{b}{bolometric diameter fit from the {\sc Nexsci} getCal tool.}
\tablenotetext{c}{since this star is a pre-main sequence star, extra care is needed when using it as an interferometric calibrator due to possible photometric and diameter variability as well as extended NIR flux; however, it appeared unresolved in our observations, as well as in those of  \citet{2005ApJ...635.1173A_KI}}
\tablecomments{Calibrator stellar diameters significantly smaller than 0.5~mas, are unresolved by the KI. 
The statistical errors given here for the bolometric diameter fit to a black body likely underestimate systematic errors of the NIR diameter of stars, but even 0.2~mas uncertainties in the diameter would not change the visibility calibration.}
\end{center}
\end{table}

The Keck Interferometer (KI) atop Mauna Kea combines the light of the two 10-meter Keck telescopes and has a baseline of 85 meters, oriented 38$^\circ$ east of north \citep{2004SPIE.5491..454C, 2004SPIE.5491.1678W}. 
We used the KI in the $V^2$ continuum mode. All data shown here are from the white-light channel of the beam combiner, which illuminates one pixel with the full $K'$-band (2-2.4~$\mu$m) to maximize sensitivity.  

The observations were conducted on the nights of Mar.~17, and Dec.~15,~2008 (UT). Details of the observations appear in Table~\ref{tab:4}. 
KI data are provided to the observer in a semi-raw state. 
The technical calibration, such as detector bias corrections, and some averaging has been applied by a pipeline reduction. The result are raw fringe contrast (visibility) measurements, which still need to be calibrated for the so-called system visibility (the visibility transfer function), and the ratio correction, which corrects for systematic flux biases between the telescopes.
Both the system visibility and the flux ratios proved to be very stable over the two nights, indicating a reliable data calibration.
Due to these stable conditions and good seeing, we did not follow up each target with a calibrator immediately, but rather alternated two targets with one visibility calibrator measurement. 
We used several different calibrator sources throughout the night to enable cross calibration, and to match the targets in brightness within $\pm~1\,{\rm mag}$ to minimize the impact of the known flux-dependence of the KI system visibility. 
{  Flux bias calibration of the visibilities was applied, which has been shown to push the systematic $V^2$ calibration errors from $\sim 0.05$ down to $\lesssim \, 0.03$~\footnote{see  {\tt http://nexsci.caltech.edu/software/KISupport/dataMemos/fluxbias.pdf}} }. 

The calibrators were selected using either the getCal planning tool \footnote{Distributed by the NASA Exoplanet Science Institute (NExScI): {\tt http://nexsciweb.ipac.caltech.edu/gcWeb/gcWeb.jsp}} or browsing {\sc SIMBAD} directly within a radius of $\sim15^{\circ}$ around the science targets. 
We only used stars from the Hipparcos and Tycho catalogs to assure high coordinate precision.
To ensure that all calibrator stars are unresolved with respect to the projected baseline,
their photospheric diameters were estimated by fitting black-body SEDs to published photometry using NExScI's fbol routine (Table~\ref{tab:3}). The fitted black-body model was compared in particular to the photometry at wavelengths longer than 2~${\rm \mu m}$ to check for dust excess that would be indicative of extended structure. 

Our data reduction followed the standard procedures developed and suggested by the NExScI team, and we made ample use of their wb/nbCalib-software suite and the respective documentation.
 The individual transfer functions, derived from  each calibrator measurement, are defined as the ratio of the measured raw squared visibility to the respective ideal uniform disk squared visibility \footnote{following the measurement principles of the KI, the measurement is linear in the squared visibility modulus}.
Then, a time and sky-location dependent average system transfer function is calculated, which picks for each target the calibrators closer than 15$^\circ$ and observed within two hours before or after the target measurement. The calibrators closer in time and space get a higher weight in the averaging process. We use the default time-weighting options. 
This averaging approach potentially minimizes the effect of a single bad calibration measurement on the data calibration, in contrast to using only the two calibrators immediately taken before and after the science measurement. This calculation of an average transfer function, based on all calibrator measurements, is particularly suitable for nights with stable observing conditions. Finally the raw data are divided by the transfer function to calculate the calibrated visibilities.

We use the standard deviation of the raw measurements as a first estimate of the uncertainties of each data point.
The resulting statistical visibility uncertainties of the individual measurements are $0.005-0.01$ in most cases, smaller than the canonical value of 0.03 (see the NExScI KI support websites), which includes margin for systematics such as slightly different observing conditions and Strehl between the calibrator and science measurements. The calibrated visibility measurements appear in Table~\ref{tab:4}.
Although the observing conditions were very good and stable throughout the night, and the calibration uncertainty seems to be slightly better, we assume the standard $\delta\,V^2\,=\,0.03$ for the analysis in this paper. To get the highest precision, it is recommended to bracket a science object with two calibrator measurements at similar flux levels which would give a slightly better time-sampling of the system visibility than ours. However, since our goal was to detect visibility changes of 0.2 over the observed range in hour angles, our conclusions do not depend on the precise uncertainty adopted.  Higher observing efficiency (i.e. time sampling) of the targets is more critical for our project than highest calibration precision, to sample as large a range of binary parameters as possible. 
Still, the stable conditions and good visibility precision helped us to achieve the sensitivity to detect circumstellar material on scales as small as 1 milliarcsecond. Fig.~\ref{fig:1} shows the {\it u,v}-plane coverage of \object{DM Tau}. This is typical for the whole sample of targets. 

\section{Results}
\label{sec:5}

\begin{figure}
\epsscale{.95}
\plotone{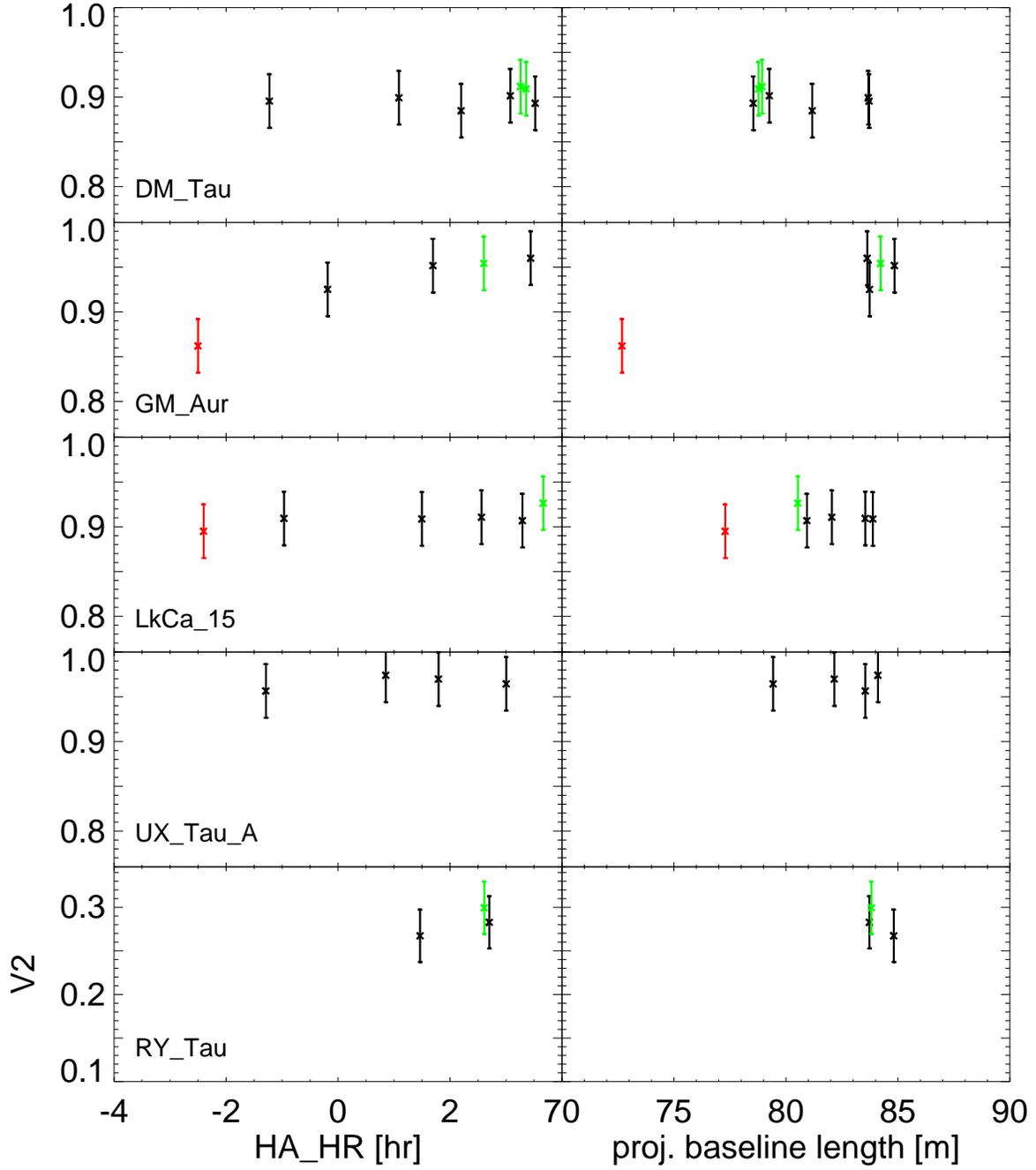}
\caption{\label{fig:2}
Calibrated $V^2$-data from our KI observation in March (green) and December (black) 2009, with 0.03 errors as discussed in the text. Note the different scaling of the ordinate axis. The red data points in the plots of GM~Aur and LkCa~15 indicate the previously published KI data from \citet{2005ApJ...635.1173A_KI}, the PTI data points for RY~Tau are not shown.}
\end{figure}

The calibrated visibilities appear in Fig.~\ref{fig:2}, plotted as a function of the hour angle and the projected baseline length.
 Since we show broad band visibilities, observed at a fixed central wavelength of about 2.2~$\mu$m, the projected baseline length scales the interferometric resolution ($\lambda/2B$) and the spatial frequencies ($B/ \lambda$) directly for all data shown.
We added previously published KI data points for GM~Aur and LkCa~15. 
For \object{RY~Tau}, we also included in our analysis visibility measurements from a previous observation with the Palomar Testbed Interferometer \citep[PTI,][]{2005ApJ...622..440A_PTI}. 
Due to the different baselines of the PTI measurements (see footnote ($d$) in Table~\ref{tab:4}), in particular different position angles at similar baseline lengths,  we do not show those data together with our KI data in Fig.~\ref{fig:2}. They probe different spatial frequencies at the hour angles and baseline-lengths shown in the figure.
The data in Fig.~\ref{fig:2} are shown as the measured squared visibility amplitudes. 

{  The photospheric stellar diameters of the observed program stars ($<~0.3~{\rm mas}$, Table~\ref{tab:1}) cannot be resolved by the KI.
However, we measure mean calibrated visibilities
ranging from $0.28\pm0.03$ for RY Tau to $0.96\pm0.03$ for UX~Tau~A.}
Such visibilities below unity rule out that our targets are compact at 2~$\mu$m (apart for the marginally resolved UX~Tau~A).
Thus, we can test our visibility data for each target against two contrary scenarios: 
\begin{enumerate}
\item  A binary companion is present, responsible both for clearing out the optically thick dust from the
inner region of the disk, and for producing some or all of the observed near infrared excesses. 
In the following section, we demonstrate that a large range of  binary parameters in the range of dynamical interaction between the binary and the inner disk edge can be excluded on basis of our KI data. 
\item As an alternative to the binary scenario, we evaluate a simple model of a disk gap, simulating the inner dust disk by a face-on circumstellar ring which contributes to the $K$-band flux. 
\end{enumerate}
The diameter estimates of such a pragmatic model approach give order-of-magnitude constraints on the location and extension of the observed emission.

As seen in Fig.~\ref{fig:2}, we mostly sample a range of  similar projected baseline lengths (75~--~85~m). 
Therefore, we cannot firmly conclude from the visibility data alone whether we resolve or over-resolve the dust emission. 
While  resolved emission would result in decreasing visibilities with longer baselines, an over-resolved emission would show constant visibilities below unity. 
More information, for instance about the $K$-band excess from the SED models, is needed to interpret the data best. 
Here, we aim at understanding at an order of magnitude level the physical location of the extended emission. Thus, details like scattered light contributions have not been considered in our analysis \citep{2008ApJ...673L..63P}.

\subsection{Constraints on the presence of Binary Companions}
\label{sec:51}

\begin{figure}
\epsscale{0.99}
\plotone{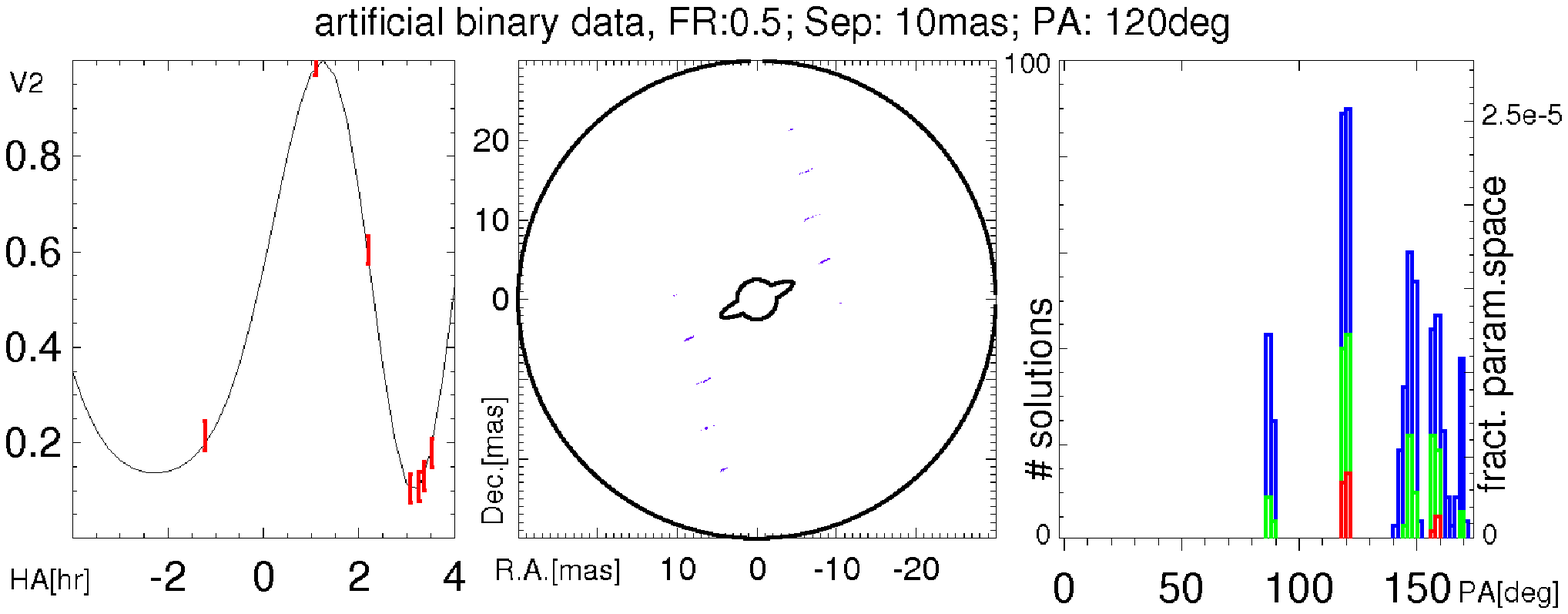}
\plotone{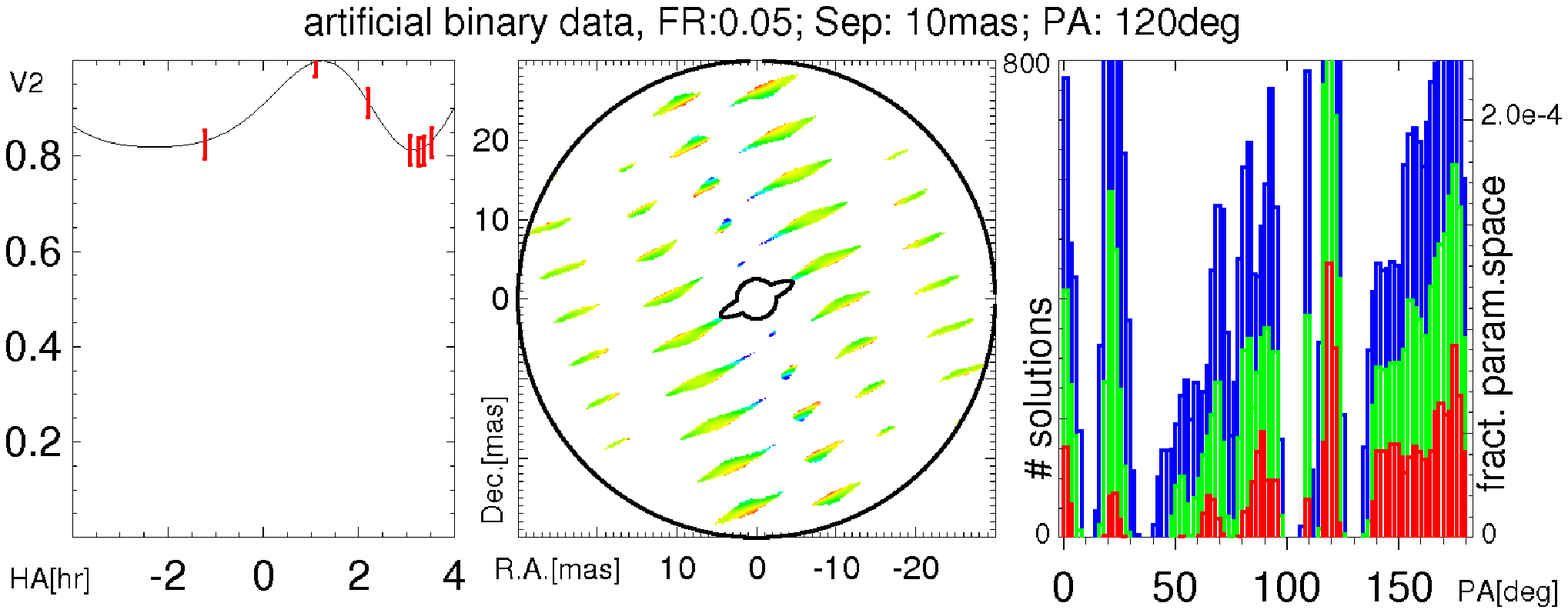}
\caption{\label{fig:80} {  Analysis of mock datasets of binary systems to evaluate the sensitivity of our analysis. The binary parameters are given on top of the figure. The left panels show the theoretical visibility curve, and the mock data points. The central panels show the positions where a companion cannot be rejected at the 99.7~\% confidence level. The color encodes the maximum flux ratio allowed at this position, see Fig.\ref{fig:81} for the color-to-flux ratio correspondence.
The right panels show histograms of the position angles of the allowed resolutions.  Here, the color encodes the confidence levels of rejection: blue (99.7~\%), green (95~\%), red (67~\%). Both the absolute number of solutions, and the respective fraction of the probed parameter space are given.}
}
\end{figure}

\begin{figure}
\epsscale{0.8}
\figurenum{4a}
\plotone{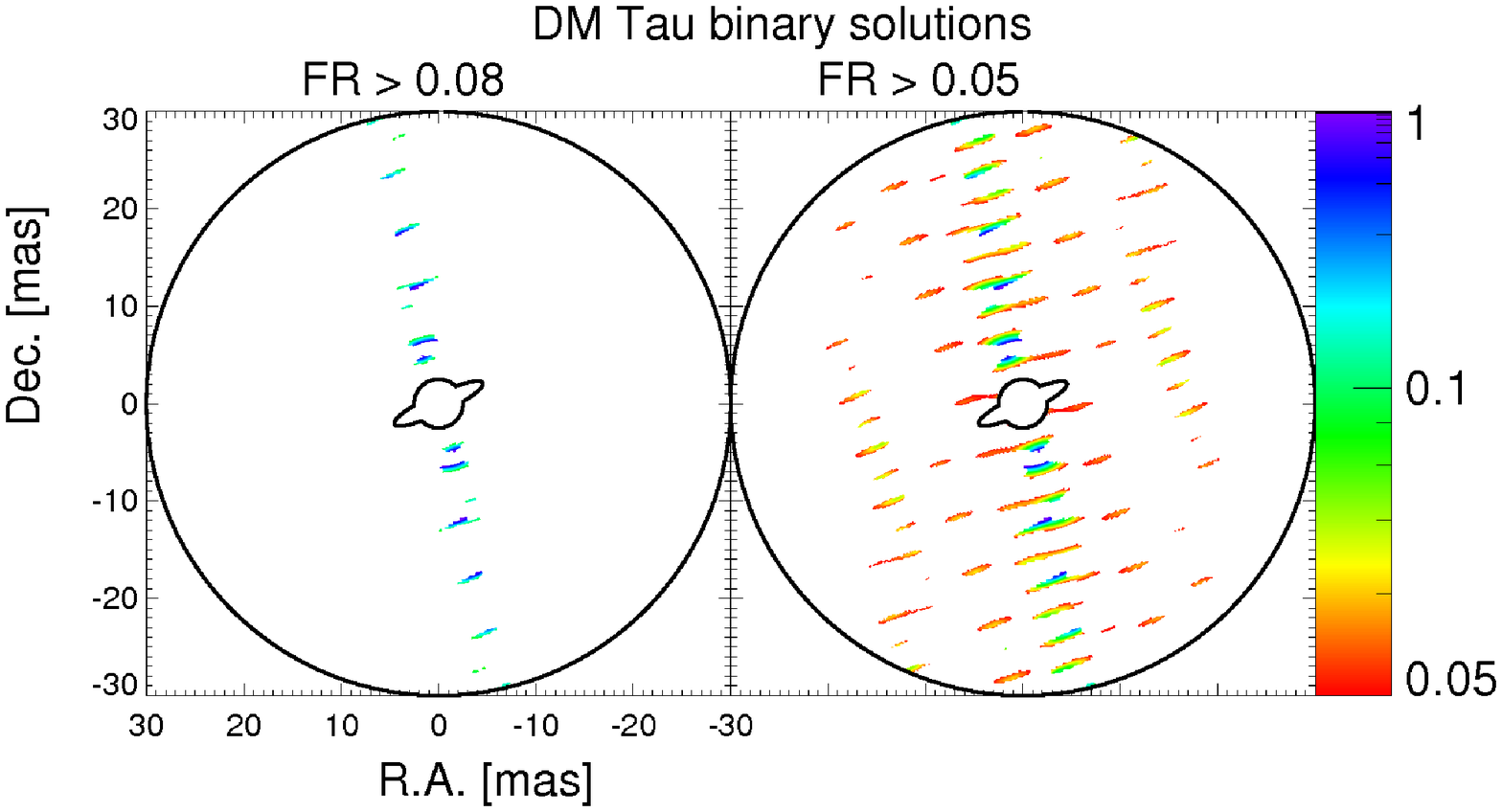}
\plotone{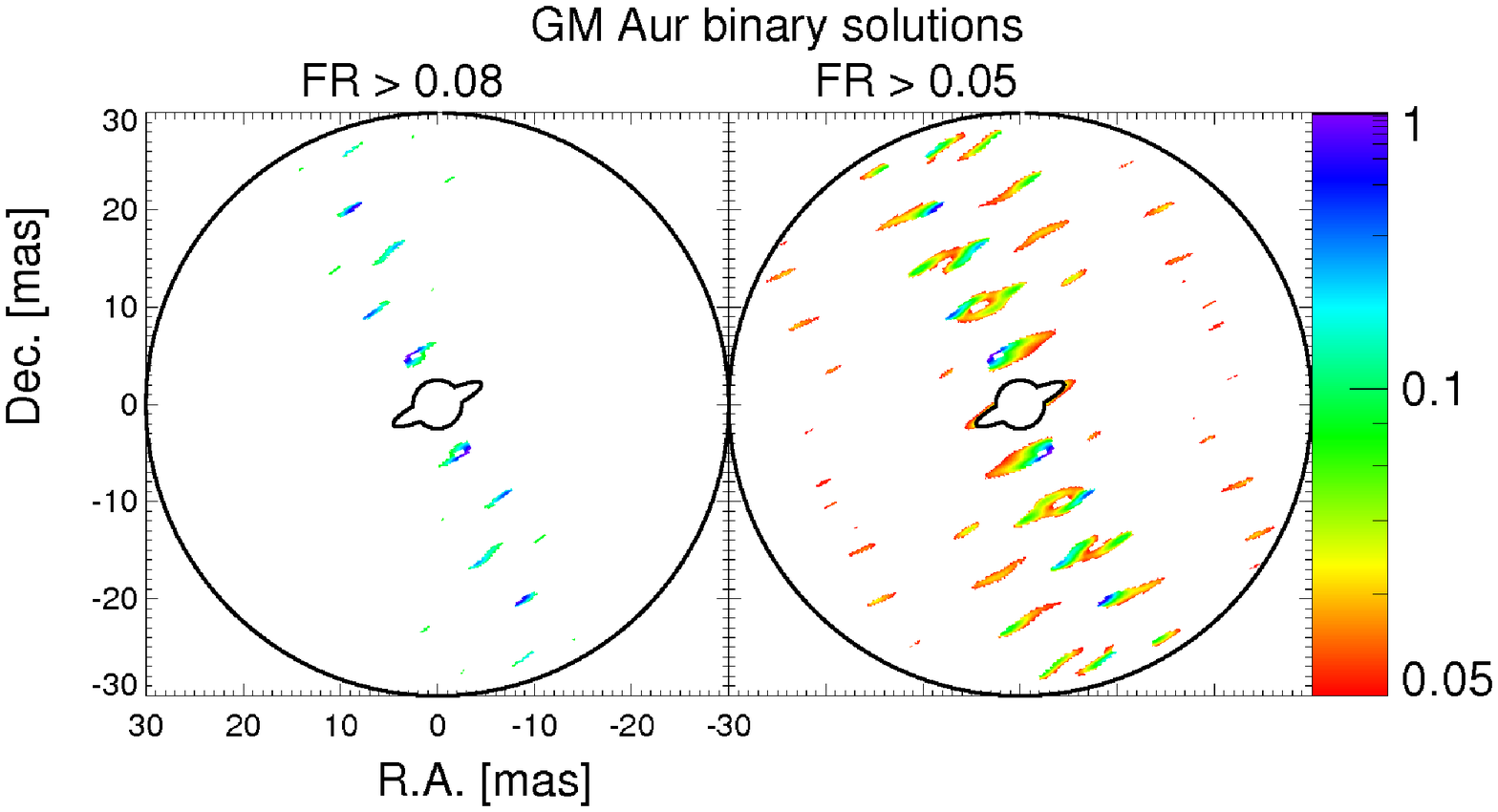}
\caption{\label{fig:81} {  Binary models of DM~Tau and GM~Aur, which are not rejected by the KI data at the 99.7~\% confidence limit. Left (right) panels the solutions down to $FR\,>\,0.08$ (0.05). North is up, and East is to the left. An inner and outer ring (at 2.5~mas and 30~mas radius) show the FoV of our search. The color bar indicates the probed flux ratio range.} 
}
\end{figure}

\begin{figure}
\epsscale{0.8}
\figurenum{4b}
\plotone{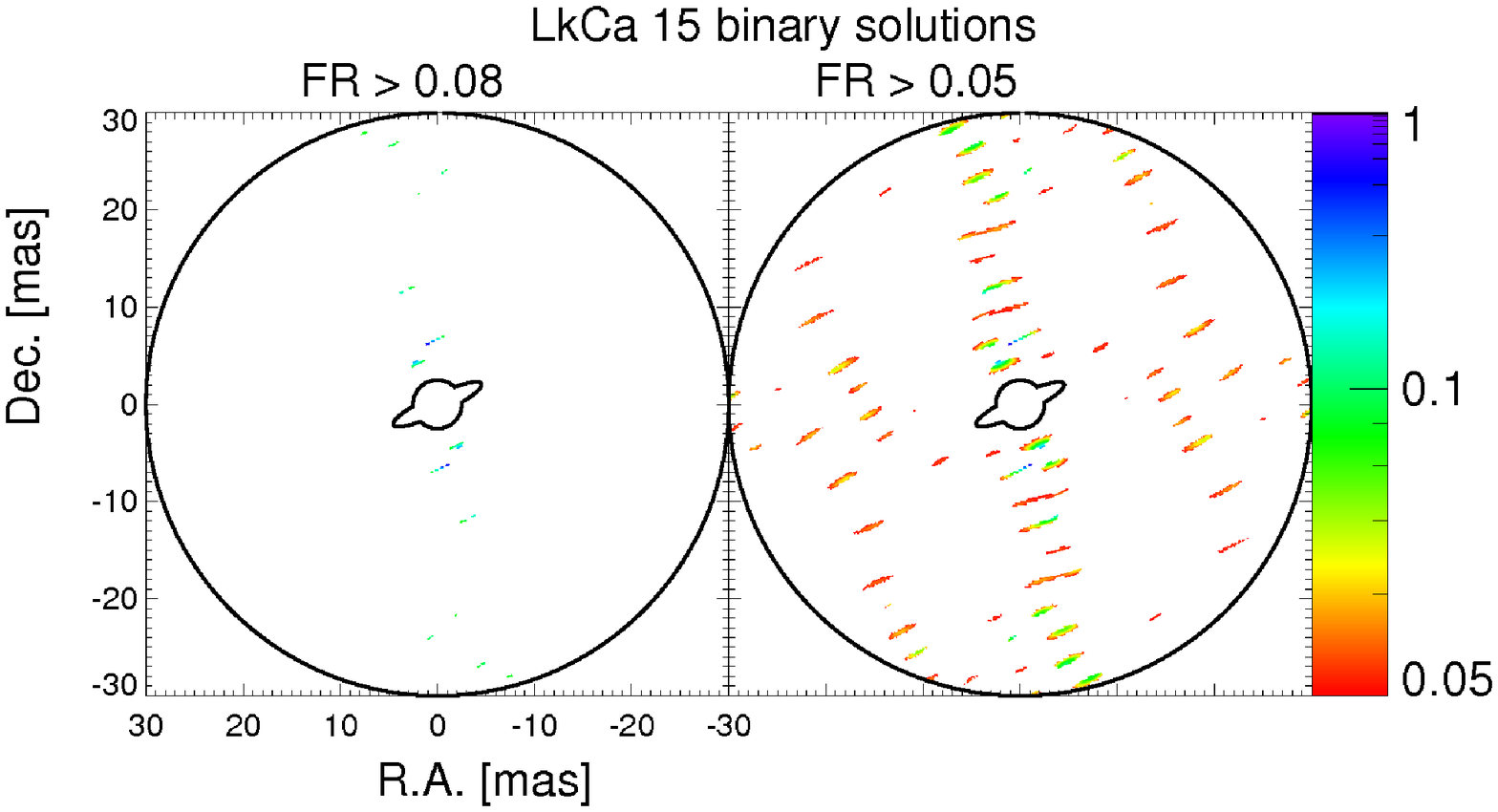}
\plotone{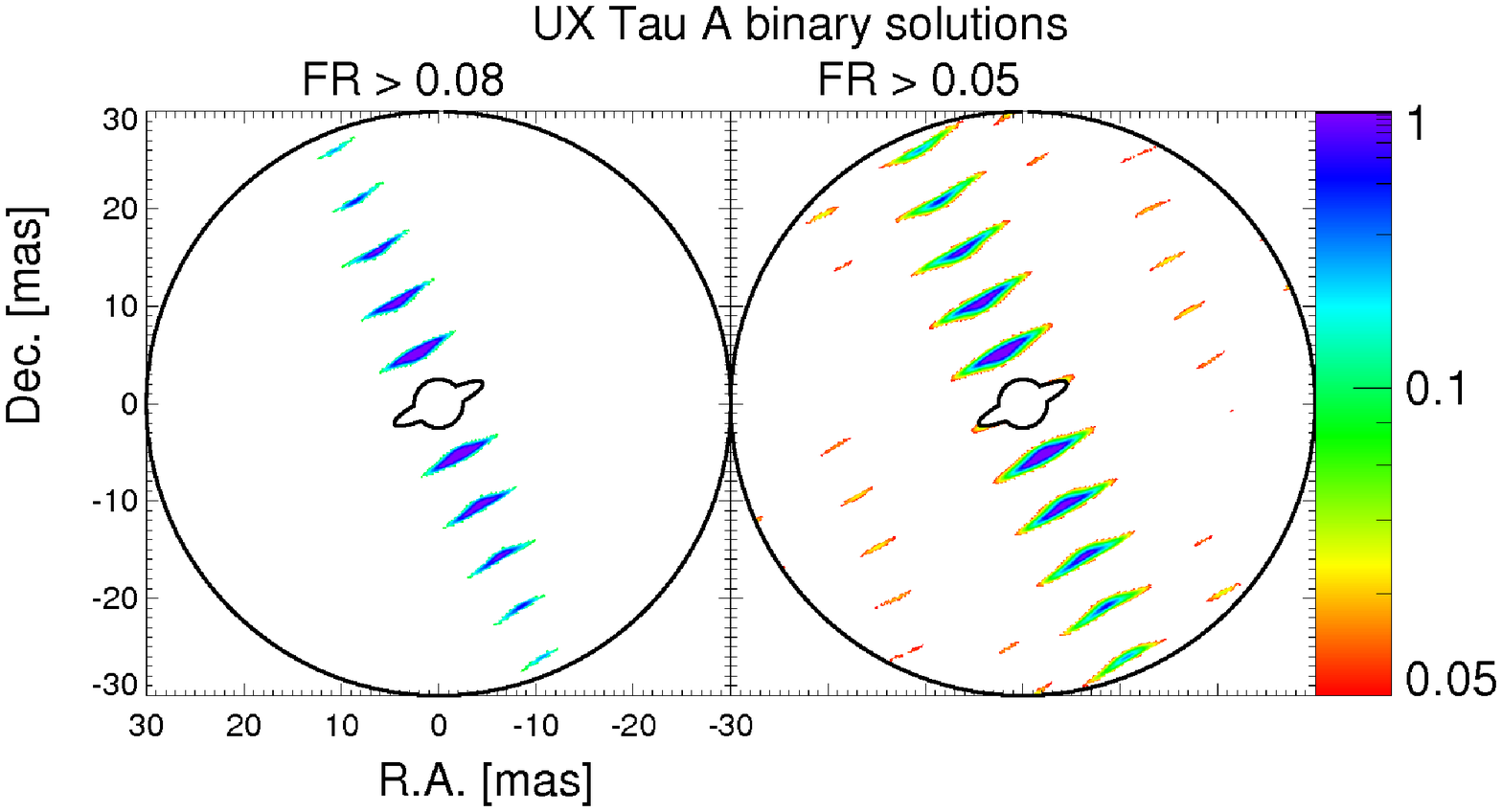}
\caption{\label{fig:881} {  Binary models for LkCa~15 and UX~Tau~A. See caption of Fig.~\ref{fig:81} for details.} 
}
\end{figure}

\begin{figure}
\epsscale{0.99}
\figurenum{5}
\plotone{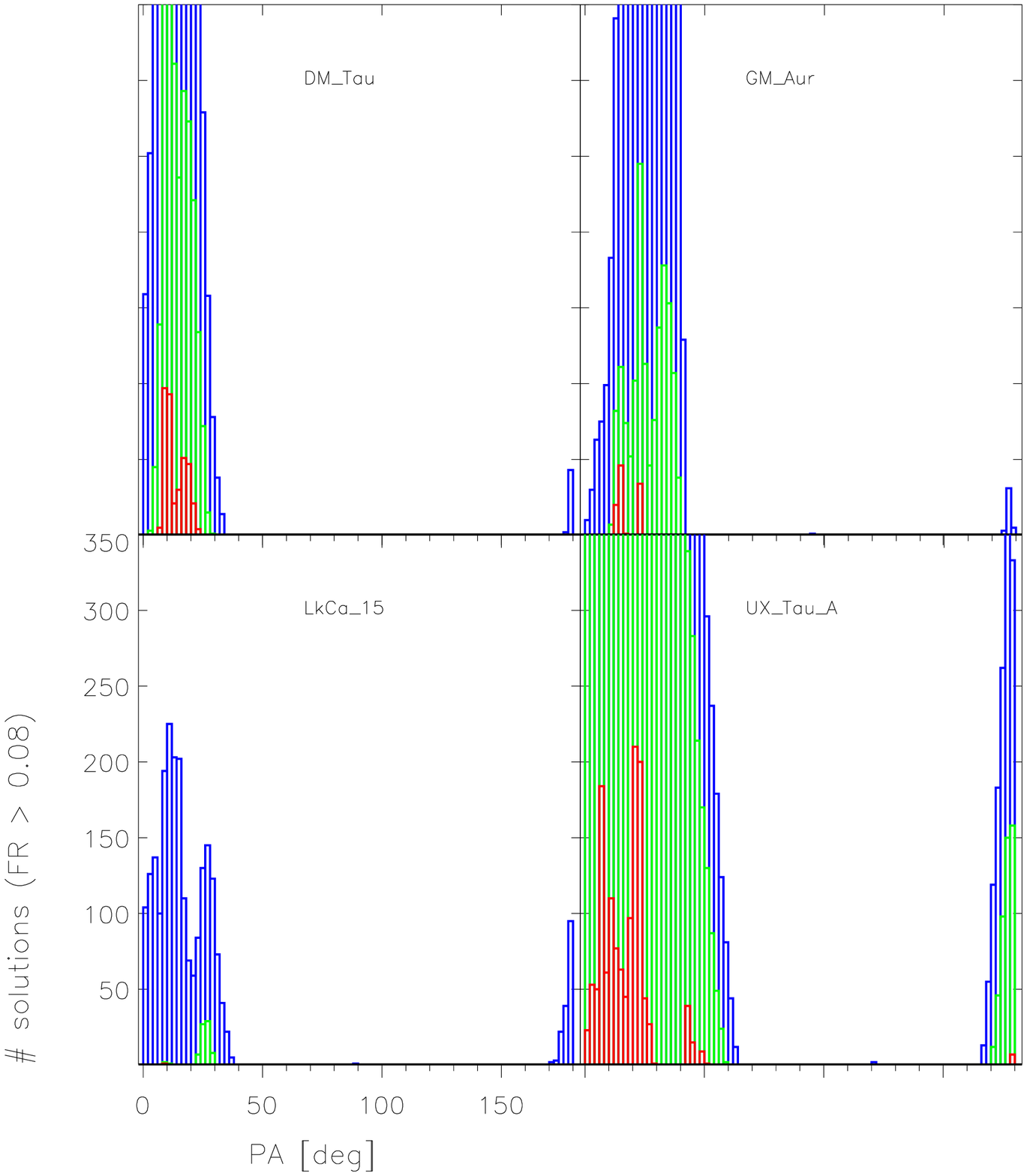}
\caption{\label{fig:82} {  Position angle histograms of the found binary solutions down to $FR\,>\,0.08$. The color encodes the confidence levels of rejection: blue (99.7~\%), green (95~\%), red (67~\%). The fixed vertical axis scaling was chosen to show reliably a few tens of solutions, which is the minimum solution density for possible binary configurations at our parameter sampling (see the upper right histogram in Fig.~\ref{fig:80}). }
}
\end{figure}
\begin{figure}
\figurenum{6}
\epsscale{0.99}
\plotone{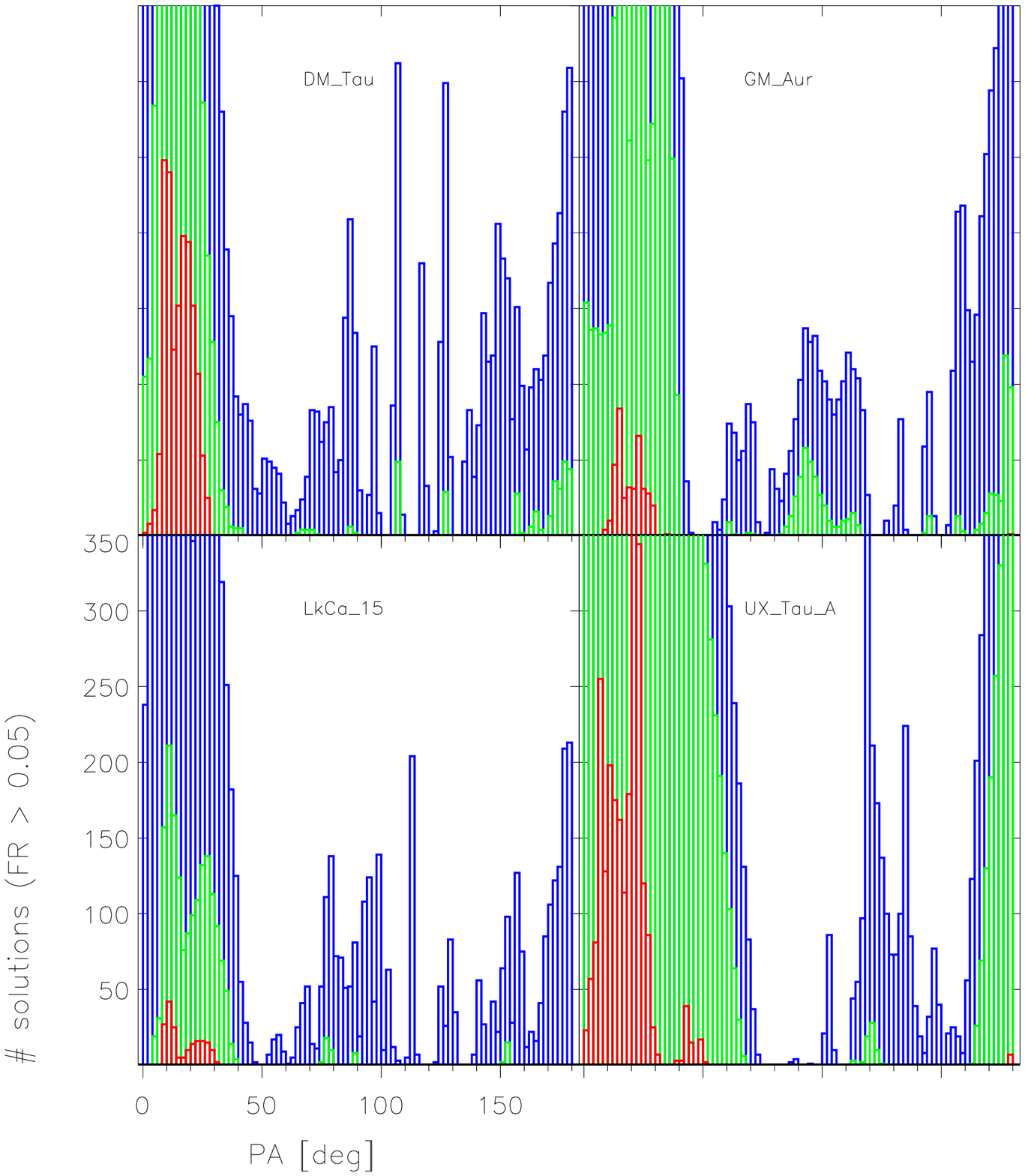}
\caption{\label{fig:83} {  Position angle histograms of the found binary solutions down to $FR\,>\,0.05$. The color coding and scaling equals Fig.~\ref{fig:82}.}
}
\end{figure}

Our overall approach is to test binary models against the observed visibilities, in order to estimate the parts
of the parameter space that are incompatible with the data.  Due to the high maximum elevation of the Taurus-Auriga star
forming region at Mauna Kea, the single baseline of the KI delivers sufficient {\it u,v}-coverage to probe a large range
of binary parameters. 
We calculated theoretical visibilities for all position angles ($PA$) from
$0\,-\,180^\circ$ (the visibility sensitivity is point-symmetric), star-star separations $\rho$ between  2.5 and 30~mas, and flux ratios ($FR$) between 1
and 0.05 times the brightness of the primary. 
For each model the reduced $\chi^2$ deviation to the data was calculated. 
Because the number of
observed data points differ for each source (and thus the number of degrees of freedom for our model
also varies) we calculate individually for each source the $\chi^2$ level that corresponds to a
formal 99.7~\% confidence level for the relevant number of degrees of freedom, based on the cumulative $\chi^2$-distribution of random measurements.
The 99.7~\% confidence limit is arbitrarily chosen to match the 3~$\sigma$ confidence level of a normal probability density function.    
Models
whose (reduced) $\chi^2$ exceeds this threshold have a probability of less than 0.3\% to be consistent with the data and are rejected \footnote{In general,
statements in this paper about a model being 'consistent' or 'ruled out' by the data refer to this
confidence level unless otherwise specified. This $\chi^2$ modeling assumes errors that are statistically independent, when in fact
calibration systematics result in correlated errors, slightly modifying the derived
probability levels. We do not attempt to model this in detail, but our choice of a
conservative estimate of 0.03 for the errors including  such calibration systematics implies that true rejection confidence
levels are most likely more stringent than the formal 99.7\% level.}. 
The thresholds are reported in Table~\ref{tab:2}.

We chose 360 and 120 linearly spaced steps for PA and $\rho$, respectively. $FR$ is sampled in 80 logarithmic steps. Thus, we examine about 3.5 million possible binary configurations per star.
{  These steps are small enough to adequately sample the parameter space and find a number of solutions if a binary would have been observed. This is demonstrated by analysing mock datasets. Fig.~\ref{fig:80} shows artificial visibility data at the $u,v$-coverage of the DM~Tau observation. The data points are distributed around the theoretical visibility curve (solid line) using a Gaussian noise with a FWHM of 0.03, matching our measurement precision. This exercise shows several properties of our analysis. In case of flux ratios close to one the sensitiviy is very high, and only a few tens of solutions are found. The sparse $u,v$-coverage results in artificial solutions which are evenly separated by multiples of the KI-resolution ($\sim$~2.7~mas), and along the position angle of the KI-baseline. Each of the colored patches in the central panels of the figure show where a companion cannot be rejected at the 99.7~\% confidence limit. The solution histogram in the right panels shows how the number of solutions depend on the confidence level of the rejection criterium. The $PA$ of the mock binary datasets is with 120~$^\circ$ chosen to be orthogonal to the KI baseline. The KI observations are most sensitive to these $PA$ due to the rotation of the projected baseline, thus this $PA$ is best suited to test the sampling of the binary parameter search grid. A decent number of a few tens of solutions is found. For binary $PA$ more aligned to the KI baseline, the number of solutions would increase.}

{  The combined PTI and KI dataset of RY~Tau, the target with the largest NIR-excess flux, excludes the complete binary parameter space probed. The other four visibility datasets allow for a few solutions, which are visualized in Fig.~\ref{fig:81}\&b. We argue in the following that these solutions are most likely artefacts of the sparse $u,v$-coverage of the observations, missing data from additional baselines as in the case of RY~Tau.}
{  In the left panels of the figure, we show all allowed solutions down to $FR\,=\,0.08$. The field of view (FoV) of our binary analysis is the 2.5~-~30~mas radius annulur patch of sky around each central star, marked in the plots}. The slight extension of the inner boundary of the FoV, orthogonal to the KI-baseline, reflects the shape and binary sensitivity of the point spread function of visibility datasets, based on one baseline. 
At each point in the FoV, we plot
color-coded the {\it brightest} companion flux ratio allowed for a companion at that location. For instance, the blue spots in the plot of DM~Tau indicate where we found binary solutions with a flux ratio close to unity.

{  The bright companion solutions are all grouped around a $PA$ of about $20\,\pm\,20~^\circ$. This is quantified in the position angle histograms of the solutions in Fig.~\ref{fig:82}. At a confidence level of 67~\%, this is true down to a $FR$ of 0.05 (Fig.~\ref{fig:83}). This $FR$ of 0.05 is the sensitivity limit of our observations, because now binary solutions start to appear all over the FoV (right panel of Fig.~\ref{fig:81}\&b). On the other hand, for $FR$ brighter than 0.08, we typically rule out more than 99~\% of the probed FoV.}
In Table \ref{tab:2}, we report the fraction of the FoV without binary solutions.
{  Although we cannot rule out completely the probed  binary parameter range for DM~Tau, GM~Aur, LkCa~15, and UX~Tau~A individually, we
can strongly rule out the possibility that the majority of our target stars are binaries. For instance, if two (three) of these stars are binaries, the
probability that the two (three) binaries would have position angles aligned with $20\,\pm\,20\,^\circ$ is about 3~\% (0.4~\%) only \footnote{Due to the point-symmetry of visibility amplitude data, these probabilities calculate as $(40/180)^N\,*\,(5-N)/5$ where N is the number of binaries out of our five targets, and (5-N)/5 is the probability to chose one of the single stars for the PTI observations.
}.
This statistical argument is further strengthened by a similar baseline-dependent constraint of the solutions to separations of multiples of the KI resolution only. Thus, at the 99.7~\% (67~\%) confidence level of rejecting binary models, it is very unlikely that more than one out of our five targets have a binary companion of 0.08 (0.05) brightness of the primary.}
The masses that such flux limits represent depend on
both the luminosity of the primary and the assumed age of the system.
Based on the PMS evolutionary tracks of \citet{1999ApJ...525..772P} and an age of  1~Myr, flux ratios of 0.1 would refer to companion masses of about 1/4 of the primary mass. 

The fact that we can rule out binaries for the major part of the tested FoV, and for a brightness ratio down to 0.05 is mostly related to the measured small {\it differential} visibility variations. In contrast, the {\it absolute} level constrains binary parameters to a lesser extent only, but is rather linked to the amount of (over-)resolved emission. An equal brightness binary would result in large ($\gtrsim\,0.2$) squared visibility variations over the probed range of spatial frequencies of the KI baseline. This amplitude of the visibility variation decreases with decreasing companion brightness (see the left panels in Fig.~\ref{fig:80}). Our binary search is limited to $FR\,\gtrsim\,0.05$ because even lower $FR$ would result in differential visibility variations compatible with models of a single resolved, extended emission structure, as discussed in the next section. Fitting binary models to such data of low differential visibility variation is very ambiguous. The visibilities of GM Aur show a statistically significant differential visibility signal of $\sim\,0.1$, different from the other four targets, but our analysis shows that this still small change in visibility does not appreciably alter the range of allowed binary solutions. We discuss this visibility trend further below in Sect.~\ref{sec:111}.

The set of binary parameters, excluded by our data, does not significantly depend on the fact that so far we probed the data against simple, dust-free binary models only. It is possible that some extended, circumbinary dust emission has been resolved out by the interferometer, resulting in the visibilities below unity. This raises the concern that the putative binary signature in the visibility variation is not recovered by the analysis because the absolute visibility level has changed with respect to the calculated one, due to the missing over-resolved dust emission in the correlated flux. 
To probe if this scenario would have an effect on the findings of the above binary analysis, we simulate the contribution of over-resolved dust emission by dividing the measured visibilities by their maximum. Then, the same binary analysis can be applied to the modified data set. The results are very similar. Binarity of RY~Tau is still ruled out within the searched parameter range, due to the large $u,v$-coverage of the combined KI \& PTI data. The modified visibilities of DM~Tau, GM~Aur, and LkCa~15 are now all close to unity, and thus are similar to the originally measured data of UX~Tau~A. Therefore the same argument holds as above. Binary solutions along the KI baseline position angle cannot be ruled out completely, but it is very improbable that all targets {  harbor} a similarly oriented binary.

Summarizing, our observations 
\begin{itemize} 
\item {  are sensitive to binary companions with flux ratios comparable to the near-IR excess fraction observed for these sources}
\item exclude almost all of the probed binary parameter range, down to companion flux ratios of 0.05 and 2.5~mas (3.5~AU) separations 
\item reject stellar binarity as the dominant mechanism in creating a transitional disk appearance
\end{itemize}

{  \citet{2008ApJ...678L..59I} searched with $K$-band Keck aperture masking interferometry for inner binaries in DM~Tau, GM~Aur, LkCa~15, and UX~Tau~A, and excluded companions with mass ratios $>\, 0.1$ over $20\,-\,160\,{\rm mas}$ separation range. Thus, we extend this search at similar sensitivity down to 2.5~mas due to the superior resolution of the KI baseline.}

\subsection{Modeling the disk emission}

For all targets but DM~Tau, previous SED and high resolution studies concluded that significantly
more than 5\% of the $K$-band flux is not emitted by the central stellar object but rather is expected to come from
some kind of circumstellar material (see also our
individual target discussions below, and the references therein).  
{  In the following, we focus on analyzing the geometric constraints derived from the interferometric data. We do not aim at differentiating between candidates for this emission. Therefore, we simply speak of hot dust as emitter, keeping in mind that other flux contributions from circumstellar gas and scattering of stellar flux are possible as well.}

As we have established in the previous section, we
can exclude nearly all possible explanations which invoke a binary companion with flux ratio $FR > 0.05$.
Furthermore, the fact that we resolve all targets with the interferometer indicates a significant
$K$-band flux contribution from extended hot circumstellar matter. 
Thus it appears to be rather
unlikely that the inner regions of these transitional disks are totally clear of dust. 
Instead, a gap must separate the innermost hot dust (visible at $2~\mu {\rm m}$) from the cooler dust which
dominates the SED at wavelengths longer than about 8~$\mu {\rm m}$. This correlates well with the
fact that all our targets have been classified as {\it classical} T\,Tauri stars, with indications
for ongoing mass accretion, which requires mass to be located close to the star.


If there {\it is} circumstellar dust radiating at 2~$\mu {\rm m}$, the expected 3d-morphology is a ring at distances where the dust grain equilibrium temperature is comparable to sublimation temperatures of 1000-1500~$K$. 
Since our observations are limited to the $K$-band (at about 2~$\mu$m), we are not sensitive to cooler dust {  outside of a few AU}. 

The limited {\it u,v}-coverage inhibits a model-free interpretation of the visibility data, but without many additional assumptions, we can evaluate the following simple disk scenario, chosen to minimize the number of model parameters.
We fit to our data a central point source plus a face-on, centro-symmetric, narrow ring, leaving only two degrees of freedom: the flux ratio between star and ring, and the ring radius. 
If, in fact, the simple model is reasonably close to the astrophysical reality, the fitted radii would represent the approximate, order of magnitude, location of the radiating material.  
Only if the emission morphology departed significantly from circular symmetry (jet-like, edge-on disk), the radii derived here would be less meaningful.
Fitting more realistic {\it inclined} rings or disks would add at least two more parameters (position and inclination angles), and would essentially require measurements of an orthogonal baseline to obtain good constraints. 
Previous measurements do not show a very large inclination ($>\,60^\circ$) for any of the targets (Table~\ref{tab:1}), so in keeping with our small {\it u,v}-coverage, we do not explore this scenario. 
We did test a three-parameter model which lets the radial thickness of the bright ring also vary, but we found no significant improvement over the two-parameter model described here. 

The results are shown in Fig.~\ref{fig:61}-\ref{fig:65}. For each target, the left panel shows the reduced $\chi^2$ of the given model parameters; red contours indicate the $\chi^2$ contour corresponding to 99.7\% confidence, as described above. For GM~Aur, LkCa~15, UX~Tau~A, and RY~Tau, we have also overplotted the disk-to-star flux ratio and disk inner radius estimates based on previous SED fits (Table 1). The horizontal dashed green lines in each plot represent a typical error $\pm 5$\% for such photometry based NIR-excess estimation. Similarly, we over-plotted for DM~Tau an upper limit of 5\%$\,K_{\rm excess}/K_{\rm total}$.
\citet{2005ApJ...630L.185C} find for GM~Aur dust inside 5~AU, and the horizontal dashed lines represent the respective dust excess. But they cannot further constrain the location of this inner dust. 
Appropriately, no vertical line is given in the figure for this target.

Qualitatively the 'allowed' regions of the $\chi^2$-plots in the left panel with $\chi^2$ close to unity show two features: a diagonal feature for small ring radii where we truly have the angular resolution {\it and} sensitivity to resolve the ring, and a wavy horizontal feature at larger radii, where the ring is {\it over}-resolved. The flux ratio of this feature is connected to the average visibility level, and refers to how much of the total flux is over-resolved.  

In the right panel of Fig.~\ref{fig:61}\&\ref{fig:65}, we plot the minimum $\chi^2$ per ring radius, from the parameter range within the SED-model estimated flux ratios (horizontal dashed lines in the left panels, also Table~\ref{tab:1}). The fact, that these minimum $\chi^2$ are often significantly below unity, does not indicate that our used conservative $V^2$ uncertainty estimates including instrumental and observational systematics, are too large for the presented data. 
However, it indicates that these systematic biases do not necessarily change from data point to data point, and that the measurements are not completely independent at this accuracy level. 
We then derive our model radius for GM~Aur, LkCa~15, and RY~Tau as the radius of the minimum $\chi ^2$, for the disk-to-star flux ratios derived from the SEDs. For DM~Tau and UX~Tau~A, we derive a lower, and a upper limit for the radius, respectively, as discussed below. Table \ref{tab:2} lists the best-fit model radius and $\chi^2$ for each target, derived using the constraint of the disk-to-star flux ratios from SEDs. The vertical dashed lines in the right panels of  Fig.~\ref{fig:61}\&\ref{fig:65} show again the radii where dust emission is expected, as derived from SED models. There is a reasonable coincidence with the results from our visibility analysis for all targets.

Details for the individual targets are given in the following sections, but here we summarize the findings qualitatively.
\begin{itemize}
\item Our data typically require a minimum flux contribution from the inner disk. 
Since this is $K$-band flux, it cannot come from very large radii, where cooler dust resides (peaking at longer wavelengths). 
Thus, our data reject the scenario of a totally cleared inner disk. Scenarii for transitional disks, with some hot dust very close to the star  and / or a transition zone with no or optically thin dust emission inside of  the cooler, outer optically thick dust appear consistent with the data. 
\item Our two-parameter star/ring model fits the data well within our confidence limits, but the radius is not well constrained by the imaging information from our KI data alone.  Unambiguous model radii can be derived if we add the $K$-band excess flux  the $K$-band excess flux estimation to the visibility analysis.
{  These radii are comparable to the 0.1~AU size scales of the innermost dust, inferred from state-of-the-art SED disk models.}
\end{itemize}

{  As discussed in Sect.\ref{sec:115}, DM~Tau might be the only target without dust such close to the star. }

\subsection{Discussion of Individual Targets}

If not cited otherwise, the background information  discussed in this section, in particular photometry and the Spitzer spectroscopy,  is published or
referenced in the survey articles of \citet[][2009]{2006ApJS..165..568F}. 
All our five targets show the
typical mid-IR excess at wavelengths longer than about 8~$\mu {\rm m}$, and Spitzer IRS spectra
reveal the 10~$\mu {\rm m}$ silicate emission feature, generated by optically thin dust in the  surface layer of optically thick disks, or in gaps void of optically thick dust. 
The strength of the silicate feature varies between targets, and in particular UX~Tau~A shows very weak silicate
emission, potentially due to larger grains in the inner disk from grain growth during disk
evolution. 

We discuss the stars below in increasing order of the expected disk contribution to the $K$-band flux as derived from SED-models.

\subsubsection{DM~Tau, expected $K_{\rm inner\,dust}/K_{\rm total}$: $\lesssim\,5\,\%$} 
\label{sec:115}
\begin{figure}
\figurenum{7a}
\epsscale{0.9}
\plotone{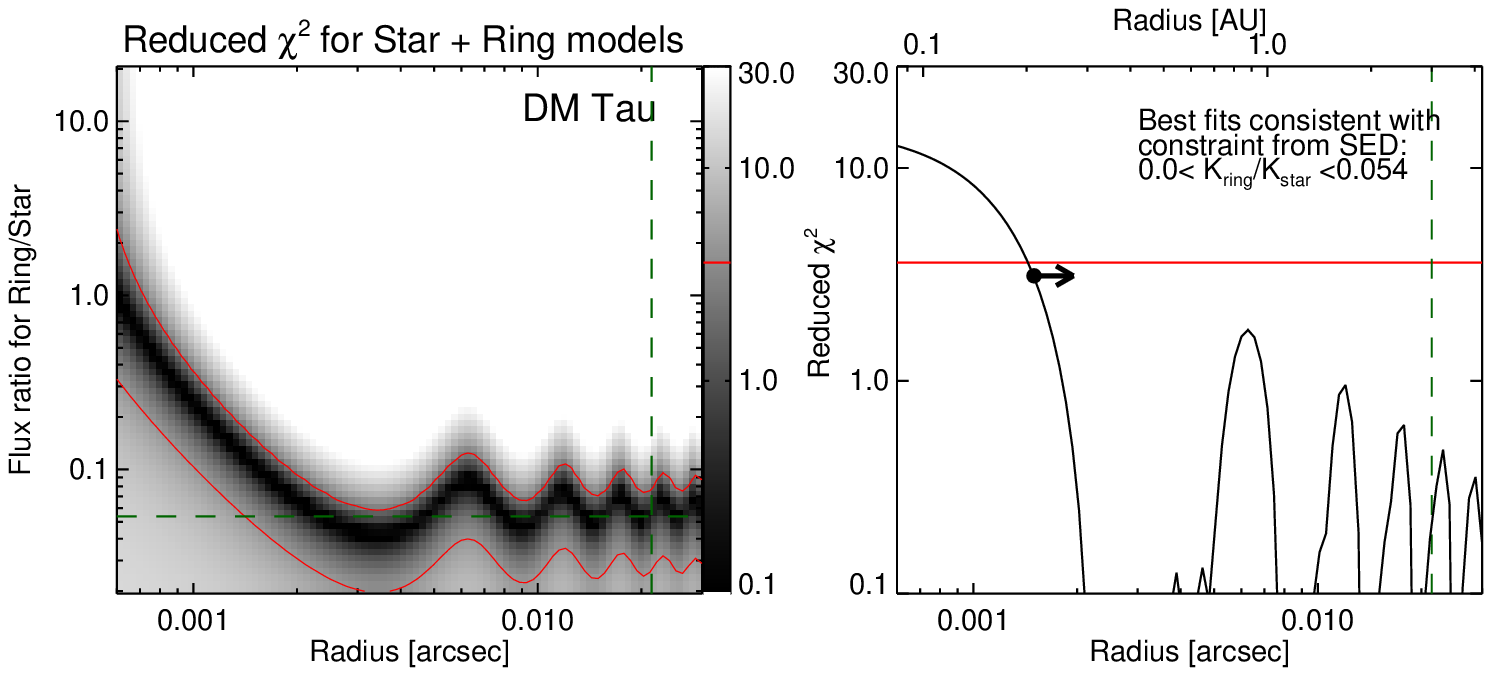}
\plotone{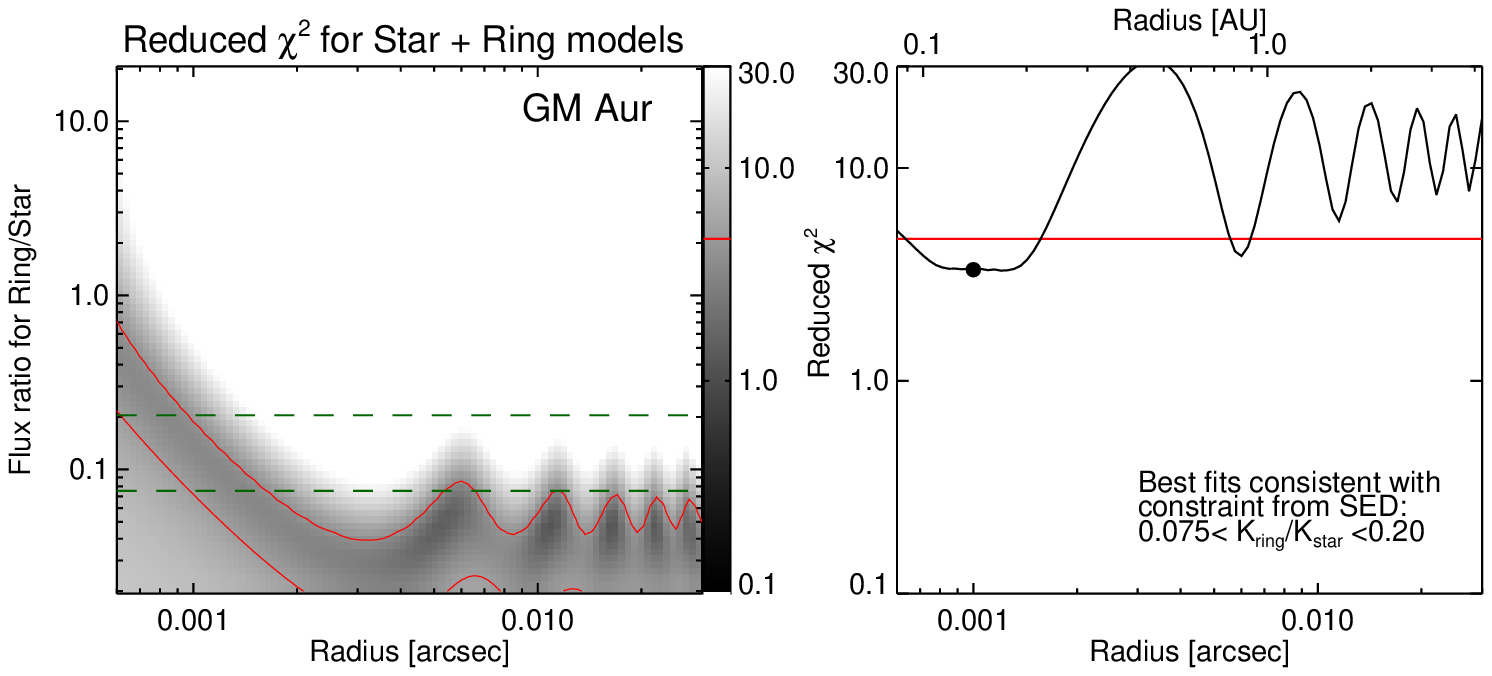}
\caption{\label{fig:61}
Star plus ring fits to the interferometric data of DM~Tau, and GM Aur. The left panels show the reduced $\chi^2$ for each set of model parameters. Red lines  in both panels trace the 99.7\% confidence contours; models outside these contours do \textit{not} fit our data, while models within these contours are possible based on the visibilities alone. The green dashed lines indicate the flux ratio and inner radius range inferred from previous SED fits (see Table 1).  The color scaling for the left panel is indicated adjacent to the right panel. 
The right panel shows the minimum $\chi^2$ per ring radius, estimated whithin the SED-derived flux ratio range. The ring diameter favored by our data is overplotted, and given in Table~\ref{tab:2}. Note that the arrow in the upper right plot of DM~Tau, indicating the lower limit, matches the 1.5~mas given in the Table. The slight discrepancy to the 99.7~\% level, the formal rejection criterion, is due to rounding. 
}
\end{figure}

\begin{figure}
\figurenum{7b}
\epsscale{0.9}
\plotone{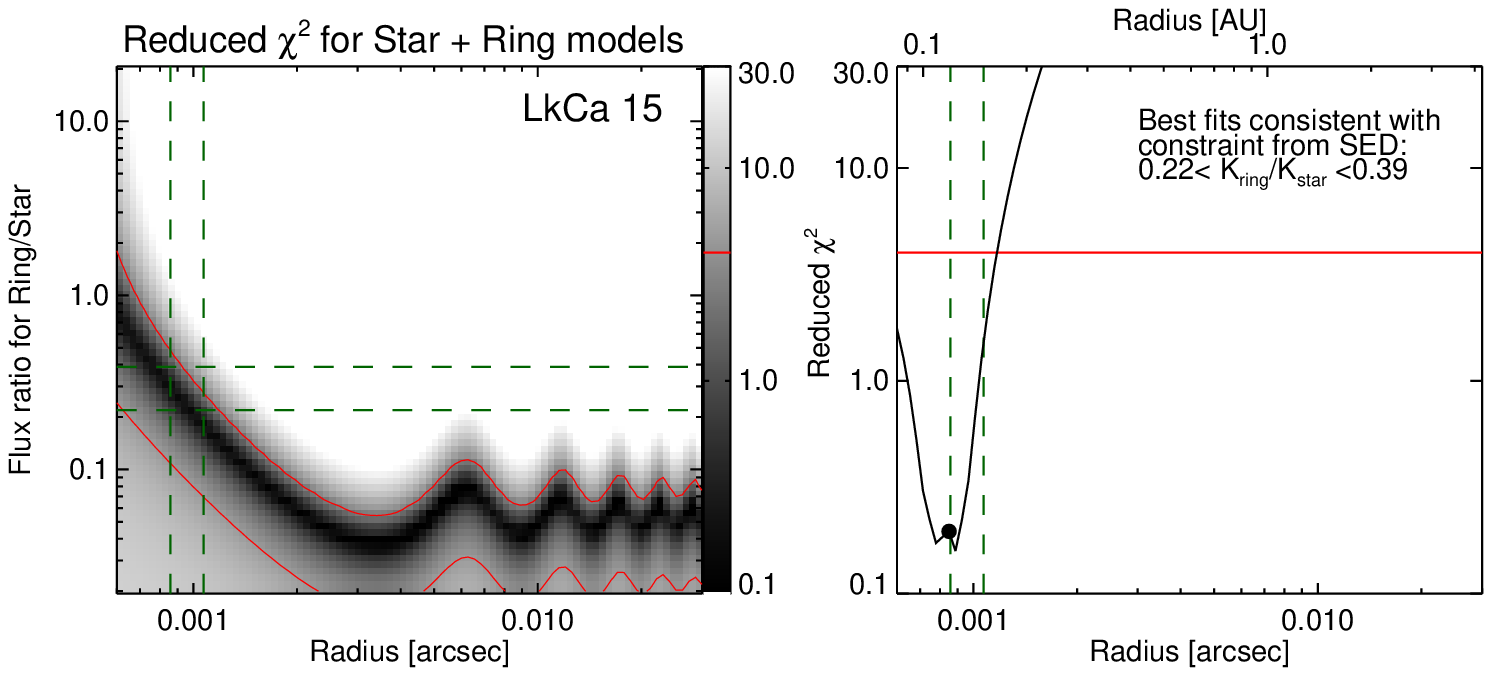}
\plotone{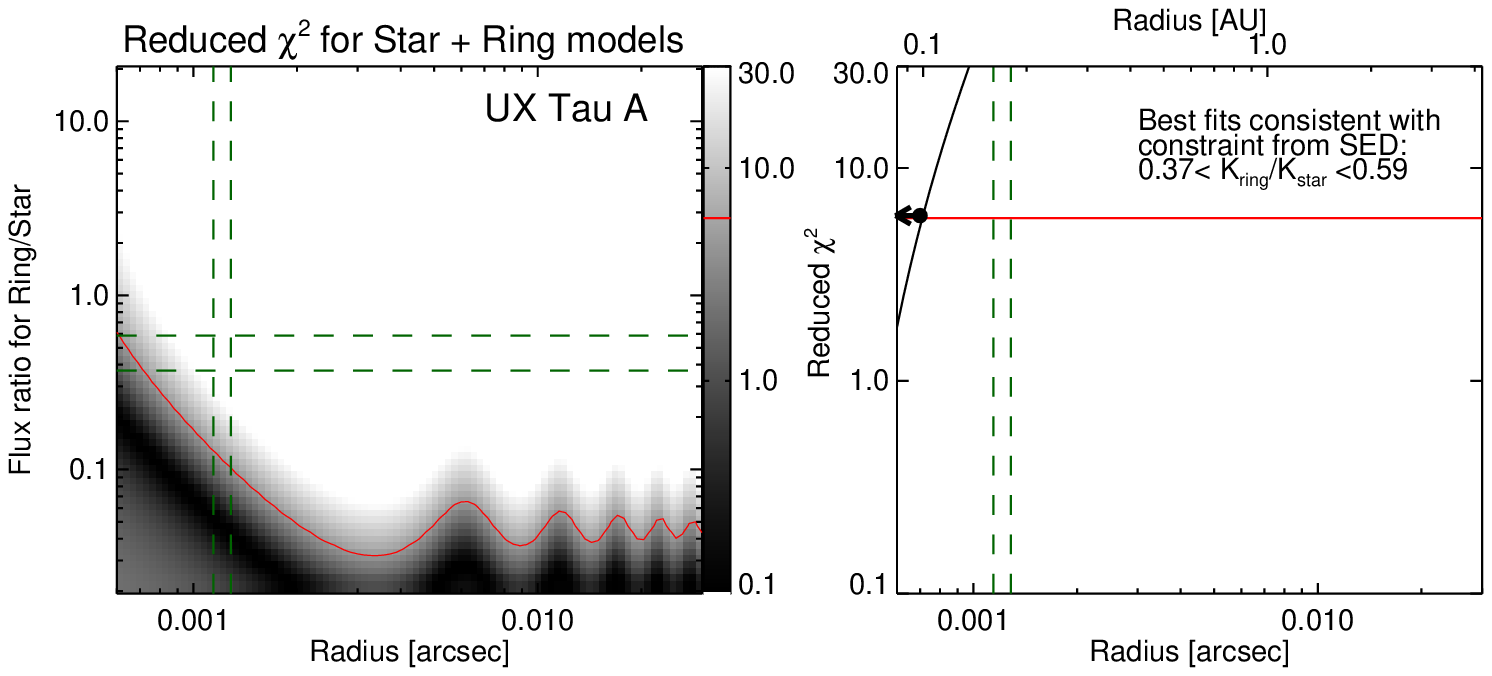}
\plotone{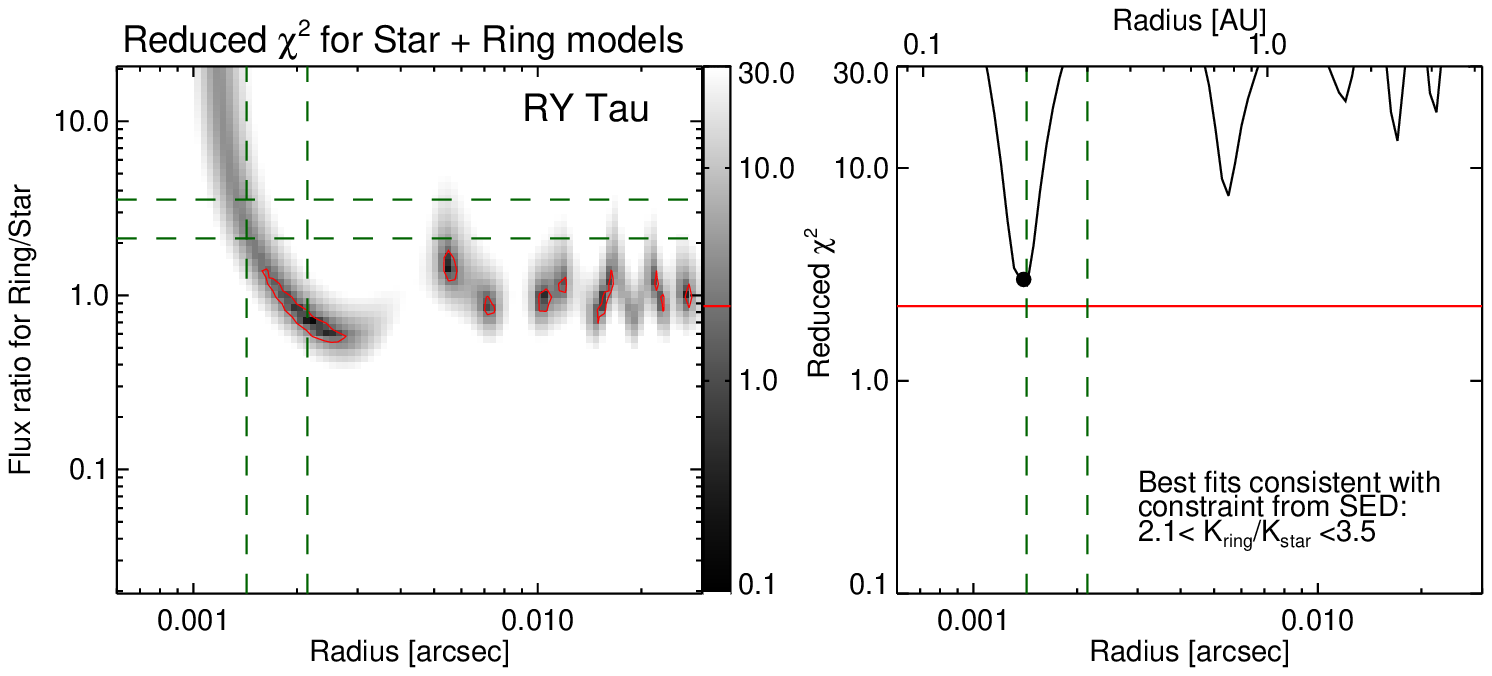}
\caption{\label{fig:65}
Star plus ring fits to the interferometric data of LkHa~15, UX~Tau~A, and RY~Tau. See caption to Figure \ref{fig:61} for a description of the plotted quantities.
}
\end{figure}

DM~Tau does not show any significant emission excess at wavelengths shorter than 8~$\mu {\rm m}$. \citet{2005ApJ...630L.185C} fit a typical dust disk model to the optical to mid-IR SED, including an IRS spectrum. They find no significant excess from dust closer than 3~AU to the star, although their model allows for small amounts of optically thin dust in the inner region. {  Their estimation of the upper limit for the optically thin dust mass is 2$\,\cdot10^{-11}\,M_{\rm \odot}$. This resembles an upper limit on the disk excess of about 5~\%.} However, our visibility measurements clearly show that at least about 5\% of the $K$-band flux is not compact photosheric emission (see the horizontal feature of good solutions along the 5\% line in Fig.~\ref{fig:61}). 
Without an a priori star-to-extended flux ratio, we cannot completely constrain the average stellocentric location of this inner emission, as we do for the other targets. But, assuming a 5\% upper limit for $K_{\rm excess}/K_{\rm total}$, we can {\it exclude} radii smaller than 0.2~AU, based on our ring-model. Given this lower limit, and the small $K$-band excess, the extended emission in DM~Tau might be scattered star light alone, without hot dust emission from very small stellocentric radii.

{  Our measurement demonstrates that even a single calibrated KI data point is more sensitive to a small contribution from extended flux at the few percent level than  spatially unresolved SED fits. }
{  A recent spectro-photometric study finds  neither a significant continuum excess  nor circumstellar CO emission at 5~$\mu$m in DM~Tau \citep{2009ApJ...699..330S}. }
The fact that there is very little 2-5$\,\mu$m circumstellar excess emission  might be due to low dust masses and an increased grain size.
{  However,  \citet{2009ApJ...699..330S} confirm the detection of mass accretion onto the star, traced by HI recombination lines. 
}

\subsubsection{\label{sec:111}GM~Aur, expected $K_{\rm inner\,dust}/K_{\rm total}$: 12\%}

%

\citet{2005ApJ...630L.185C} fitted the same type of model to GM~Aur as was used for DM~Tau, using optically thin inner and optically thick outer dust. They find for GM~Aur a dust disk distribution which is not completely devoid of inner dust: the inner, optically thin dust appears to not extend to radii larger then 5~AU, and the outer optically thick dust is outside of about 24~AU. Thus, they inferred a gap between the inner and outer dust distributions. 
With their given flux ratio, the KI data constrains the emission location of the bulk of the inner dust to be much further in, at radii of $\sim 1$ mas, (0.15AU). 
\citet{2005ApJ...635.1173A_KI} report a slightly larger ring radius fitted to their single KI-data point ($1.58\pm0.6$ mas), which is fully consistent with our findings due to our now increased {\it u,v}-coverage.

Due to the simplicity of the face-on ring model, the resulting radius is an order-of-magnitude estimation, but the resolution advantage of the interferometer over SED models alone remains apparent. 
The visibility data constrains the location of the $K$-band excess emission significantly better.
We did observe a significant trend of increasing visibilities with increasing hour angle and baseline length (Fig.~\ref{fig:2}). This could indicate that we observe an inclined disk, in which case our fit radius is a lower limit to the true radius. 
In particular, the published data point from \citep{2005ApJ...635.1173A_KI} shows a {\it lower} visibility, i.e. a larger size, at a shorter baseline, i.e. at {\it lower} angular resolution. This cannot be explained with a circular-symmetric structure. Indeed, \citet{2000ApJ...545.1034S} find with mm interferometry a disk inclination of 56 degrees for the cool outer disk around GM Aur. This (outer) disk inclination is confirmed by {\sc Nicmos} images of scattered light \citep{2003AJ....125.1467S}. 
The circular ring model hits with a disk inclination of 60~$^\circ$ a validity limit. 
Even higher inclinations could lead to a significant under estimation of the disk size, if modeled by a circular ring, but our sample does not include stars with disk inclinations beyond this limit.

Our limited {\it u,v}-coverage does not allow us to reliably fit an inclined disk / ring model, but a broad range of inclined disk models are consistent with the observed visibility trend.
A combined mm/NIR modeling approach might be worthwhile but bears some caveats. The inner hot disk might have a different orientation than the cool outer disk, and it is also possible that we observe a very close binary system instead of an inner disk, outside the fit range of Sect.~\ref{sec:51}. 

\subsubsection{LkCa~15, expected $K_{\rm inner\,dust}/K_{\rm total}$: 23\%}


Although its visibilities are lower than for GM~Aur, we fit a smaller ring radius of about 0.85 mas to the data of LkCa~15, due to the larger flux contribution of the inner disk. \citet{2000ApJ...545.1034S} fit similar inclination and position angles to the cool mm-disk of LkCa~15 and GM~Aur.  Given the increased disk dominance, a trend of the visibility versus hour angle should be stronger in LkCa~15, if the orientation angles are comparable. However, we do not observe such a trend within our uncertainties, which might be due to a combination of the interpretation caveats given in the previous section.  

Based on near infrared spectroscopy, \citet{2008ApJ...682L.125E_LkCa} found that LkCa 15 has a gap between inner and outer optically thick dust, placing it as a member of the so-called pre-transitional disks. 
Such SED-models suggest that LkCa 15's optically thick inner disk is located between 0.12 and 0.15 AU \citep{2007ApJ...670L.135E_LkUX,2008ApJ...682L.125E_LkCa}. Our high angular resolution data support these findings by confining the bulk of the inner dust radiation to within 1 mas (0.15~AU) from the star. Indeed, our best-fit ring radius is $0.85 \pm 0.05$ mas = $0.12\pm0.01$ AU. Note that the fact that our radius is slightly below the SED model estimation might be due to modelling an inclined inner disk (Table~\ref{tab:1}) with a face-on ring.

\subsubsection{UX~Tau~A, expected $K_{\rm inner\,dust}/K_{\rm total}$: 32\%}

%
UX~Tau~A's properties, as measured by the KI, differ from our other targets. It is less spatially resolved. In fact the data are marginally consistent with a point source. Also, the Spitzer IRS spectrum shows a significantly weaker 10~$\mu {\rm m}$ silicate emission feature than in any other transitional disk. 
Both properties are surprising at first glance given the relatively large inner disk contribution as derived from SED models. The proposed explanation is an optically thick inner disk containing primarily large dust grains \citep{2007ApJ...670L.135E_LkUX}. Those authors find characteristic radii for UX Tau A's optically thick inner and outer disks of 0.16~AU and 56~AU respectively. 
At the given flux ratio, our modeling suggests a ring radius smaller than 0.7~mas (0.1~AU) for a face-on disk. 
This roughly agrees with the numbers from the SED models, given the order-of-magnitude quality of our face-on ring models. 
In particular, the radius solutions of our circular ring model are likely slightly too small due the 60~$^\circ$ inclination of the system  \citep{2007ApJ...670L.135E_LkUX}. 

\subsubsection{RY~Tau, expected $K_{\rm inner\,dust}/K_{\rm total}$: 73\%}

In RY~Tau, the NIR excess over the photospheric radiation is much more prominent than in the other targets. This is likely related to RY~Tau's earlier spectral type, and thus hotter photosphere. 
The higher luminosity may cause a puffed-up inner disk rim and increased NIR excess.  
Its visibilities are much lower than any other target (mean $V^2 = 0.28$), confirming that in the $K$ band, this system's light is dominated by the extended disk. For RY~Tau we took only three new KI data points, and added them to the measurements of an earlier PTI experiment \citep{2005ApJ...622..440A_PTI}. The availability of several baselines leads to the complete exclusion of the entire binary parameter range probed, as discussed in Sect.~\ref{sec:51}, assuming that the brightness distribution did not change significantly during the few years between the PTI observations and our KI experiment. 

For the sake of comparison, we show in Fig.~\ref{fig:65} our simple star plus face-on ring fit to the data, which shows the same two qualitative features as before: the diagonal feature at low separation and the horizontal feature at larger separations--marking the resolved and over-resolved ring radii. The face-on ring model fits the data fairly well, although due to the much greater number of baselines probed, the reduced $\chi^2$ significance limit is much reduced compared to our other targets, and no models for $K_{\rm inner\,dust}/K_{\rm total}$=73\% fit within our 99.7\% confidence level. This most likely suggests that we have moved beyond the applicability of our simplistic face-on ring model.
Due to the addition of the  PTI baselines, we are sensitive to the inclination of RY~Tau. \citet{2005ApJ...622..440A_PTI} estimated an inclination of $25\pm 3$ degrees for a ring model, consistent with our data. Our inferred ring radius of $\sim 1.5~{\rm mas}$ agrees with the earlier findings within the uncertainties. Note that \citet{2003ApJ...597L.149M} reports a spectro-photometric model fit for RY~Tau with a high disk inclination, which is not supported by the interferometric data \citep[see the discussion in ][]{2005ApJ...622..440A_PTI}.
This demonstrates the {  ambiguity of such model fits without including high angular resolution imaging data}. 


\subsection{Comparison with CoKu~Tau/4}
\label{sec:52}
As noted above, recent high-resolution aperture-masking by \citet{2008ApJ...678L..59I} has revealed that the supposedly transitional disk around CoKu~Tau/4 is instead a circumbinary disk around a near-equal flux binary with a (projected) star-star separation of 53~mas.
Thus, there is no need to invoke planet formation to produce its inferred disk clearing within $\sim 10$ AU. 
But for all other known and well-studied transitional disks in Taurus, we have probed the binary separation range that could be responsible for creating the inner holes implied by the target SEDs, and concluded that the interferometric data rule out binarity as the predominant cause for the lack of 
hot dust emission in these systems.  CoKu~Tau/4 appears to be the exception, not the rule.

Is there any systematic difference between CoKu~Tau/4 and other transitional disks which might 
let us distinguish between these two classes of objects based on other
criteria?
{  Table~\ref{tab:1} lists the physical properties of the CoKu~Tau/4 system as compared to our target stars. Despite of their similar age and spectral type,  CoKu~Tau/4 is about twice as bright as DM~Tau. This hints already to the existence of an equal-mass binary, as found in the diffraction limited Keck images described above, and supports at the same time our negative result of the binary search, given that our program stars do not show such stellar over-luminosity.}

\citet{2008ApJ...678L..59I} note further, that one of the main differences is that only CoKu~Tau/4 is a weak-lined TTS: its spectrum does not show significant signs of mass accretion \citep{2005ApJ...621..461D}. However all five targets of our sample are mass-accreting classical TTS, with typical accretion rates of $10^{-9..-8}\,{\rm M_{\odot}\,yr^{-1}}$ \citep{2005ApJ...630L.185C}.
This matches with the theoretical work of \citet[][and references therein]{1994ApJ...421..651A}, which suggests that accretion of a circum-binary disk onto the central stars is inhibited by the resonant torques at the circumbinary dust disk inner edge.
Thus, the binary in the CoKu~Tau/4 system may be the cause not only of the disk hole, but also for the insignifcant accretion of the disk material onto the stars. Furthermore, our data confirm the presence of hot dust in the inner regions of  transitional disks.
This suggests a scenario in which true transitional disks will in general retain some small amount of hot, radiating dust inside of the outer, partially cleared optically thick disk. They also will have signs of accretion onto the central star. 
AU-scale binaries, however, may clear disks in a way that they mimic transitional disk SEDs, but lack ongoing accretion of material on sub-AU scales, inside the binary, and also inside the outer optically thick dust, cleared by dynamical interaction with the binary.

\begin{table}
\begin{center}
\caption{\label{tab:2}
Derived binary limits and ring properties
}
\begin{tabular}{lccccc}
\tableline
\tableline
    & DM~Tau & GM~Aur & LkCa~15 & UX~Tau~A & RY~Tau \\
\tableline
\% FoV without companion $>0.05$  & 95.1 & 94.8 & 97.2 & 94.4 & 100. \\
\% FoV without companion $>0.08$  & 99.4 & 99.3 & 99.9 & 98.0 & 100. \\
\tableline
Best-fit ring radius $\rho$ (mas)& $>\,1.5$ & 1 & 0.85 & $<\,0.7$ & 1.4 \\
Best-fit ring radius $\rho$ (AU) & $>\,0.2$ & 0.14 & 0.12 & $<\,0.1$ & 0.2 \\
\tableline
$\chi_{\rm red}^2$-limit (99.7~\%) & 3.6 & 4.6 & 4.0  & 5.8 & 2.2 \\
\tableline
\end{tabular}
\tablecomments{ The first two rows give the fraction of each
field of view for which our data rule out the presence of any binary
companions brighter than the stated flux ratio relative to the primary. The field-of-view probed is the annulus from 2.5~mas~-~30~mas; inside of 2.5~mas our sensitivity
rapidly degrades and some binary solutions with flux ratios near unity are allowed. The next two rows give the best-fit ring properties derived for face-on
rings, assuming the $K$ band flux excess for each target given in Table 1. The last row gives the 99.7~\% $\chi^2$ confidence levels for each dataset, which were used for rejecting model fits. }
\end{center}
\end{table}

\section{Conclusions} 
\label{sec:6}

Within a single night of repeated observations, our experiment was sensitive to close binaries with 2.5~-~30~mas separation and flux ratios down to 0.05. We were able to rule out nearly all possible binary
companions within that parameter range. 
{  For four of our five targets, we cannot entirely rule out all
possible companions. The remaining solutions cover $\lesssim\,2\%,\,(\lesssim\,6\%)$ of the probed FoV for companion flux ratios larger then 0.08 (0.05) of the primary's brightness. Those solutions are preferentially aligned with the baseline of the Keck Interferometer, where our sparse sampling of the $u,v$-plane renders our measurements insensitive. Observations along an orthogonal baseline (e.g. from CHARA or VLTI) should suffice to completely rule out the remaining part of parameter space, as was the case for RY Tau. However, already now a simple statistical analysis reveals that it is very unlikely that more than one of our target stars would indeed harbor a binary of separations larger than the minimum binary separation probed.  These interferometric observations extend the finding of similar binary searches, based on diffraction limited 10~m class telescope imaging, down to ten times smaller projected separations.
We conclude that, unlike in the CoKu Tau/4 system, binarity is
in general not responsible for either clearing disk holes 
to produce transitional-disk-type SEDs, or for the near-infrared excess over the photospheric emission. }

Instead, we spatially resolve a fraction of the $K$-band emission in all five stars, ranging between $\sim5-70\%$ of the total $K$-band flux.
By fitting a toy disk model to the data, we find that this inner disk emission typically comes from radii of about 1~mas (0.15~AU), consistent with previous work.  
In particular, these findings are consistent with recent disk models fitting spatially unresolved spectro-photometric data of transitional disks. 
A next step for studying transitional disks at high angular resolution could be to fit more realistic models simultaneously to the observed visibilities and the SED, as has been recently done for a few systems (Tannirkulam et al. 2008, Pinte et al. 2008). 

We confirm that the transitional disk phase is often characterized by several distinct dust zones: 
an inner (of order 0.1~AU), and an outer
part (of order $>10$~AU) which are not smoothly connected by a continuous distribtion of optically thick material. 
This supports the
general hypothesis that these young objects are indeed in a transitional
evolutionary state between primordial optically thick disks and optically thin disks.
In each of our targets, the habitable zones are devoid of optically thick dust
emission.
The fact that we resolve excess emission very close to the star in a transitional disk, which lacks such emission further out,  underlines that the evolution of a primordial disk is not as simple as a clearing from the inside out due to photoevaporation.
In fact, the presence of gaps suggests that we may see the effect of planet formation on disks.

\acknowledgments

MDP was supported by a NSF Astronomy \& Astrophysics Postdoctoral Fellowship.
We are grateful to the team at WMKO and NExScI for making these observations a success. We thank R. Akeson for providing the reduced data of the previously published PTI and KI measurements of some of our sources.
The data presented herein were obtained at the W.M. Keck Observatory, which is operated as a scientific partnership among the California Institute of Technology, the University of California and the National Aeronautics and Space Administration. The Observatory was made possible by the generous financial support of the W.M. Keck Foundation. The authors wish to recognize and acknowledge the very significant cultural role and reverence that the summit of Mauna Kea has always had within the indigenous Hawaiian community.  We are most fortunate to have the opportunity to conduct observations from this mountain.
The Keck Interferometer is funded by the National Aeronautics and Space Administration as part of its Navigator program.
This work has made use of services produced by the NASA Exoplanet Science Institute at the California Institute of Technology.
This research has made use of the SIMBAD database, operated at CDS, Strasbourg, France.



{\it Facilities:} \facility{Keck:I ()}, \facility{Keck:II ()}



\clearpage
\end{document}